\documentclass[10pt, a4paper]{article}
\usepackage{graphicx} % Required for inserting images
\usepackage[british,UKenglish,USenglish,american]{babel}
\usepackage{authblk}
\usepackage{xcolor}
\usepackage[babel]{csquotes}
\usepackage{url}
\usepackage{amsmath}
\usepackage{amstext}
\usepackage{amssymb}
\usepackage{mathdots}
\usepackage{setspace}
\usepackage{stmaryrd}
\usepackage{xcolor}
\usepackage{color}
\usepackage{subfig}
\usepackage{algorithm}
\usepackage{algpseudocode}
\usepackage{comment}
\usepackage{todonotes}
\usepackage{bbm}
\usepackage{pgfplots}
\usepackage{enumitem}
\usepackage{tikz}
\usepackage{pgfplots}
\usepackage{array}
\usepackage{booktabs}
\usepackage{comment}
\usepackage{adjustbox}
\usepackage{booktabs}
\usepackage{siunitx}
\usepackage{multirow}
\usepackage{xr}
\usepackage{xr-hyper}
\usepackage{hyperref}

\usepackage{empheq}

\graphicspath{{images/}}

\usepackage{stackengine,scalerel}

\newcommand\obullet[1]{\ThisStyle{\ensurestackMath{%
  \stackon[1pt]{\SavedStyle#1}{\SavedStyle\kern.3\LMpt\bullet}}}}

\newcommand{\be}{\begin{equation}}
\newcommand{\en}{\end{equation}}

\makeatletter

\newcommand*{\addFileDependency}[1]{% argument=file name and extension
\typeout{(#1)}% latexmk will find this if $recorder=0
\@addtofilelist{#1}
\IfFileExists{#1}{}{\typeout{No file #1.}}
}\makeatother

\newcommand*{\myexternaldocument}[1]{%
\externaldocument{#1}%
\addFileDependency{#1.tex}%
\addFileDependency{#1.aux}%
}
\myexternaldocument{Suppl}

\begin{document}

\title{Modeling the prion protein-mediated transport of extracellular vesicles on the neuron surface}
\author{G.~Pozzi$^{1}$,  G.~Mazzilli$^{2}$, G.~D'Arrigo$^{3}$, C.~Verderio$^{3}$, G.~Legname$^{4}$, \\S.~Turzi$^{5}$, P.~Ciarletta$^{2}$\footnote{Corresponding author. E-mail: pasquale.ciarletta@polimi.it} \\
	$^{1}$ DISMA, Politecnico di Torino, corso Duca degli Abruzzi 24, 10129 Torino, Italy\\
    $^{2}$ MOX Laboratory, Dipartimento di Matematica, Politecnico di Milano,\\ piazza Leonardo da Vinci 32, 20133 Milano, Italy. \\
    $^{3}$ Institute of Neuroscience of Milan, CNR National Research Council of Italy, Vedano al Lambro, 20854, Italy.\\
    $^{4}$ Department of Neuroscience, Scuola Internazionale Superiore di Studi Avanzati (SISSA), Trieste, Italy.\\
    $^{5}$ Dipartimento di Matematica, Politecnico di Milano, piazza Leonardo da Vinci 32, 20133 Milano, Italy. \\
    }

\maketitle
\date{}

\begin{abstract}
Neurodegenerative diseases are among the leading causes of global mortality, characterized by the progressive deterioration of specific neuron populations, ultimately leading to cognitive decline and dementia. Extracellular vesicles (EVs) are believed to play a role in the early stages of these diseases, acting as carriers of pathogens and contributing to neuroinflammation and disease propagation. This study presents a mathematical model aimed at characterizing the movement of EVs bearing prion protein (PrP) on their surface along neuronal surfaces. The model, informed by experimental data, investigates the influence of PrP and actin polymerization on EV transport dynamics and explores the possible interplay between passive and active mechanisms.
EVs isolated from non-human astrocytes were analyzed under three conditions: untreated control (Ctrl), neurons treated with Cytochalasin D (CytoD-HN), and EVs treated with Cytochalasin D (CytoD-EV).
The mathematical model is data-driven, testing different hypotheses regarding the underlying transport mechanisms. In the CytoD-EV dataset, EV movement was modeled using a flashing Brownian ratchet to represent directed motion. For active transport in the CytoD-HN set, a symmetric periodic potential was used to describe EV rolling along the neuron surface. The Ctrl scenario incorporates both mechanisms, reflecting a more complex transport behavior.
A sensitivity analysis and comparison between numerical predictions and experimental data suggest that the model effectively captures key features of EV motion, providing a quantitative framework to interpret different transport regimes. While some variability remains, the approach offers a promising basis for future investigations into the role of cytoskeletal dynamics in EV-mediated disease propagation.
\end{abstract}

\section{Introduction}
\label{sec::Introduction}

Neurodegenerative Diseases (NDs) are among the top ten leading causes of death worldwide. Recent experimental findings have shed light on their etiology and pathogenesis \cite{alzheimer20122012, dugger2017pathology, hardy2006genetics, prusiner2013biology}. In this landscape, particular attention was devoted to the study of prion diseases. Individuals affected by such pathologies present the accumulation of a structural conformer of the cellular prion protein rich in $\beta$-sheet conformation
\cite{damberger2011cellular, sarnataro2017cell, surewicz2011prion}. Such misfolded proteins, known as prions (PrP$^\text{Sc}$), act as seeds propagating the pathogenic conformation in the central nervous system through a mechanism that resembles viral infections \cite{jucker2018propagation, stephenson2018inflammation}. This mechanism is speculated to be one of the major reasons behind the continuous progression of these diseases \cite{legname2017elucidating}. Moreover, prions are insoluble and thus aggregate with other cellular components leading to the formation of deposits, which in turn lead to the death of neurons. The most common neurodegenerative diseases, as Parkinson's and Alzheimer's Diseases, share similar characteristics with prion diseases and thus they are considered prion-like diseases \cite{frost2010prion}.

Recent studies have sparked a great deal of interest in Extracellular Vesicles (EVs) because of their involvement in the intercellular communication in both physiology \cite{holm2018extracellular} and pathology, where they partecipate in the spreading of neurodegenerative diseases \cite{hill2019extracellular, gallart2020alzheimer} through neuron-to-neuron transfer of toxic material  \cite{sardar2018alzheimer, wang2017release}.
EVs are membranous structures, which contain various molecules (e.g. ATP, actin filaments, miRNA) and carry surface receptors through which they can interact with target cells and deliver their content \cite{drago2017atp}.
They are released by different types of cells into the extracellular microenvironment.
Recent findings have reported the ability of large EVs to displace in an interconnected network of neurons, even 'jumping' from one cell to another \cite{d2021astrocytes} and to travel along the axon \cite{polanco2018exosomes,gabrielli2022microglial}, thus being potential carriers of pathogens in neurodegenerative conditions.

Although the motion of these EVs has been already described in D'Arrigo \textit{et al.} \cite{d2021astrocytes}, its mechanical characterization has been little investigated and the underlying nature of EV movement is still poorly understood. Two types of mechanisms have been hypothesized for the first time by D'Arrigo et al. \cite{d2021astrocytes} to explain the behavior of EVs. A \textit{passive} movement, that occurs when EV motion is driven by the rearrangement of the neuronal cytoskeletal network, to which the vesicle is bound by a neuronal receptor, and  an \textit{active} movement, due to the presence of actin filaments in EVs, that makes the EVs capable of actively rolling over receptors on the neuron surface.

Thus, investigating the onset and development of NDs may lead to several challenges due to the different time and length scales of the processes underlying these pathologies. 
Although several mathematical models have been proposed to describe the development of NDs at the organ level \cite{weickenmeier2018multiphysics, weickenmeier2019physics, sampaoli2022toy} and the associated damage at the cellular and sub-cellular level \cite{van2015tau,de2018physical, andrini2022mathematical, riccobelli2021active}, the mechanism of seed transport by EVs is still poorly described from a mathematical perspective. However, different approaches have been proposed so far to model the active and passive movement of generic cargoes on biological surfaces.

The simplest description of this such phenomena can be traced back to the modeling of the axonal transport \cite{BROWN20091, brown2000slow}. In a mathematical description of this mechanism, Brown \textit{et al.} assumed that it could be represented as a four state Markovian process, where the neurofilaments can switch between two persistent directional states, anterograde or retrograde, but also move or pause in either state \cite{brown2005stochastic}. However, a more sophisticated description of the slow axonal transport was proposed several years earlier by Bloom and Reed \cite{blum1989model}, who for the first time hypothesized the existence of a motor, that can provide the underlying motive force while forming reversible bonds with elements of the slow transport system. Their model thus resulted in a system of mass action laws describing the aforementioned chemical interactions.

A step towards a more mechanistic description was made in the attempt of describing the operating mode of biomolecular motors, i.e. specialized proteins (myosin V, kinesin, dynein) that are able to perform directed walks, carrying organelles or EVs from one part of the cell to another. The most popular theoretical approaches adopted in modeling these active molecular units leverage the Brownian ratchet theory \cite{ait2003brownian}, based on a variant of the Generalized Langevin Equation (GLE), proposed by Feynman \cite{feynman1971feynman}. The idea of a mechanical force induced by nonequilibrium fluctuations in an anisotropic system and capable of generating a drifted motion was reproposed by Ajdari and Prost \cite{ajdari1992drift} and Magnasco \cite{magnasco1993forced} with the flashing Brownian ratchet. Unlike the Brownian ratchet, which is a one-dimensional diffusion process that drifts towards a minimum of a periodic asymmetric saw-tooth potential, the flashing Brownian ratchet is a stochastic process that alternates between two regimes: a one-dimensional Brownian motion and a Brownian ratchet \cite{ethier2018flashing}. 
In the wake of these works, a few contributions leveraging formal analytical tools were proposed in the attempt of giving further characterization on the drifted motion emerging from flashing ratchet \cite{qian2000mathematical, peskin1994correlation}. Also, Astumian and Bier \cite{astumian1994fluctuation} rigorously demonstrated that zero-average random fluctuations of the height of the kinetic barriers of the chemical reactions or a net force can cause a Brownian particle on nonsymmetric periodic potential to move upward against a constant applied force. Later on, the coupling between the mechanics and the chemical kinetics was also discussed by Keller and Bustamante \cite{keller2000mechanochemistry}.

Thereafter, several variations on the Brownian ratchet theory were proposed. Goychuk et al. \cite{goychuk2014molecular} presented a model where the dynamics of kinesin molecular motor, which takes into account the effects of the cytosol viscoelastic nature through a power-law memory kernel, is coupled with the cargo movement by elastic linkers that make it able to fluctuate independently from the motor. Similarly, Marbach et al. in \cite{marbach2022nanocaterpillar} modeled the dynamics of nano-caterpillars equipped with high number of polymeric legs, mimicked by elastic springs, that bind and unbind to a surface in a reversible manner. Lastly, a completely different approach with respect to the Brownian ratchet theory was recently proposed by García-García \textit{et al.} \cite{garcia2019guided}, who introduced a prototypical model of a guided active drift, where a Brownian particle leverages the presence of an external guiding field through a coupling between the mechanical degrees of freedom and a chemical reaction. In their model, the active drift on the particle results from a “strong” violation of detailed balance.

The aim of the present article is to propose a mathematical model to characterize the motion of astrocyte-derived EVs on their surface with the ultimate goal of shedding lights on mechanisms involved in the propagation of misfolded protein, such as prions, in prion-like diseases. The article is organized as follows. In Section \ref{sec::methods}, we first describe the acquisition of microscopies data EVs moving on neuronal processes, testing the effect of Cytochalasin D, a drug that inhibits the actin polymerization, and we then discuss the post-processing analysis.
We then propose a mathematical model to \textit{in-silico} reproduce the experimental observations and we detail its numerical implementation. In Section \ref{sec::results}, we present the main outcomes of the data analysis extracting the qualitative and quantitative information on the EV motion. We later show the results of the numerical simulations, that are finally discussed in Section \ref{sec::discussion}, together with a few concluding remarks.

\section{Materials and methods}
\label{sec::methods}
\subsection*{Ethics statement}
All the experimental procedures to establish primary cultures followed the guidelines defined by the European legislation (Directive 2010/63/EU), and the Italian Legislation (LD no. 26/2014). 

\subsection{License statement}
Video recordings employed in this work are the same used in the paper by D’Arrigo \textit{et al.} \cite{d2021astrocytes}, and utilized for the analysis described below under the license CCBY 4.0 Deed, Attribution 4.0 International, consultable at the following link: \url{https://creativecommons.org/licenses/by/4.0/} .

\subsection*{Experimental protocol}
\label{subs::protocol}
Astrocytic cultures were established from the hippocampi and cortices of rat Sprague-Dawley pups (P2) (Charles River, Lecco, Italy), Briefly, after dissection, tissues were dissociated using trypsin (0.25\%, Gibco, Thermo Fisher, Leicestershire, UK) and DNase-I (Sigma-Aldrich, St. Louis, MO, USA) for 15 min at \ang{37}C, followed by fragmentation with a pipette. Dissociated cells were plated on poly-L-lysine-coated (Sigma Aldrich, St. Louis, MO, USA) T75 flasks in minimal essential medium (MEM, Invitrogen, Life Technologies, Carlsbad, CA, USA) supplemented with 20\% fetal bovine serum (FBS) (Gibco, Life Technologies, Carlsbad, CA, USA) and glucose (5.5 g/L, Sigma Aldrich, St. Louis, MO, USA). To obtain a pure astrocyte monolayer, microglial cells were harvested from 10–14-day-old cultures by orbital shaking for 30 min at 200 rpm. Hippocampal neurons were established from the hippocampi of 18-day-old fetal Sprague Dawley rats (E18) of either sex (Charles River, Lecco, Italy). Briefly, dissociated cells were plated onto poly-L-lysine treated coverslips and maintained in Neurobasal with 2\% B27 supplement (Invitrogen, Carlsbad, CA, USA), antibiotics, glutamine and glutamate (Sigma Aldrich, St. Louis, MO, USA). Neurons were used from 2 to 17 DIV (Day-In-Vitro). To block cytoskeleton dynamics, neurons were treated with 1 $\mu$M Cytochalasin D (Sigma Aldrich, St. Louis, MO) for 1 hour \cite{d2021astrocytes}.

EVs were isolated from about 15 million astrocytes exposed to 1 mM ATP (Sigma-Aldrich, St. Louis, MO, USA) for 30 min in 10ml of Krebs-Ringer's HEPES solution (KRH) (125 mM NaCl, 5 mM KCl, 1.2 mM MgSO4, 1.2 mM KH2PO, 2 mM CaCl2, 6 mM D-glucose, 25 mM HEPES/NaOH, pH 7.4). Conditioned KRH was collected and pre-cleared from cell debris by centrifugation at 300× g for 10 min (twice) with a refrigerated centrifuge (ALC 4227 R, rotor ALC 5690). Medium/large EVs were then pelleted from the supernatant by centrifugation at 10,000× g for 30 min with a refrigerated centrifuge (VWR CT15FE, rotor Hitachi T15A61, k-factor 409 and volume involved 2ml) and resuspended in 150 $\mu$l of neuronal medium. EVs were used immediately after isolation by adding 25 $\mu$l of the EV suspension to each coverslip of neurons, in a final volume of about 400 $\mu$l (final EVs concentration \SI{5.88e6}{EVs\per \micro l}) before optical tweezers manipulation. 
The purity and size of EVs were characterized by Western blot, Cryo-EM and Tunable resistive pulse sensing (TRPS) analysis, as described in D’Arrigo et al. Briefly, Western Blotting analysis showed that EVs were positive for the EV markers Alix and Annexin-A2 and negative for the intracellular organelles markers GS28 and TOM20. Cryo-EM indicated that EVs ranged from 20 nm to 1300 nm, and TRPS analysis revealed a mean diameter of 291.39 ± 3.58 nm (see Section \ref{SI::size_distribution} of the Supplementary Information).
In one experimental group, isolated EVs were incubated with 3 $\mu$M Cytochalasin D for 1 hour at room temperature in 150 $\mu$l of neuronal medium before being delivered by optical tweezers \cite{d2021astrocytes}. 

Large EVs were delivered on the neuron surface by an IR laser beam (1064 nm, CW) collimated into the optical path of an inverted microscope (Axiovert 200 M, Zeiss, Oberkochen, Germany). Optical trapping and manipulation of large EVs was performed following the approach previously described27. Briefly, large EVs produced by astrocytes were added to neurons plated on glass coverslips and maintained in 400 $\mu$l of neuronal medium in a 5\% CO2 and temperature-controlled recording chamber at \ang{37}C. Single large EVs were trapped and positioned on a selected neuron by moving the cell stage horizontally and the microscope lens axially. After about 30 seconds, the laser was switched off to prove EV–neuron interaction. EVs were considered adhered when they remained attached to neurons after switching off the laser. EVs were considered in motion if the displacement from the contact point is greater than the EV diameter \cite{d2021astrocytes}.

EVs were live imaged in brightfield using a digital camera (High Sensitivity USB 3.0 CMOS Camera 1280 × 1024 Global Shutter Monochrome Sensor, Thorlabs, Newton, NJ, USA) at a frame rate of 2 Hz and recorded for about 20 minutes.
Collected movies were divided into three datasets:
\begin{itemize}
    \item \textit{Ctrl dataset}: control dataset, i.e. no treatment (39 samples);
    \item \textit{CytoD-HN dataset}: only neurons are treated with CytoD (19 samples);
    \item \textit{CytoD-EV} dataset: only EVs are treated with CytoD (15 samples);
\end{itemize}
for an overall number of 73 experimental samples.
Only large EVs whose size ranges from 200 to \SI{1300}{\nano \meter} were trapped and delivered to neuron surface for a matter of visibility under the microscope. Example of reconstructed trajectories are shown in Fig.\ref{fig:traj} (top row).
In the following, with the aim of giving a mechanical description of the EV movement, we analyzed the videos of the experiments where the EVs showed a non-negligible movement.

\subsection*{Data post-processing}
\label{subs::postproc}
\subsubsection*{Tracking of EV trajectories}
The coordinates indicating the position of each EV were acquired from each of the video frames using a custom code in \texttt{MATLAB} \cite{MATLAB}. The code returned a .txt file containing a data table with the coordinates $(x, y)$ of the EV position per frame every 0.5 seconds in chronological order.

\begin{figure}
\begin{tabular}{c c c}
\centering
CytoD-HN & Ctrl & CytoD-EV \\
    \includegraphics[width=0.3\linewidth]{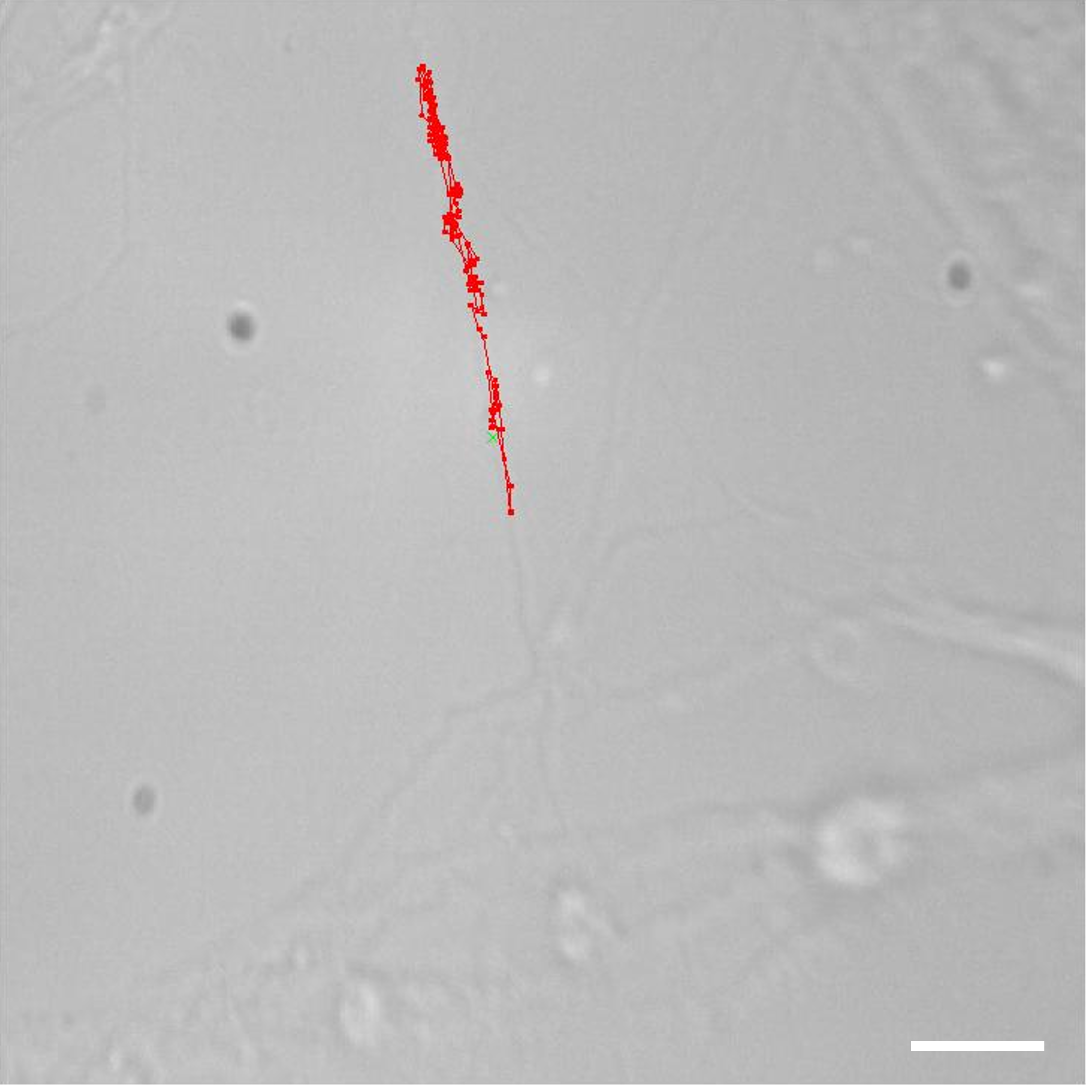} 
   & \includegraphics[width=0.3\linewidth]{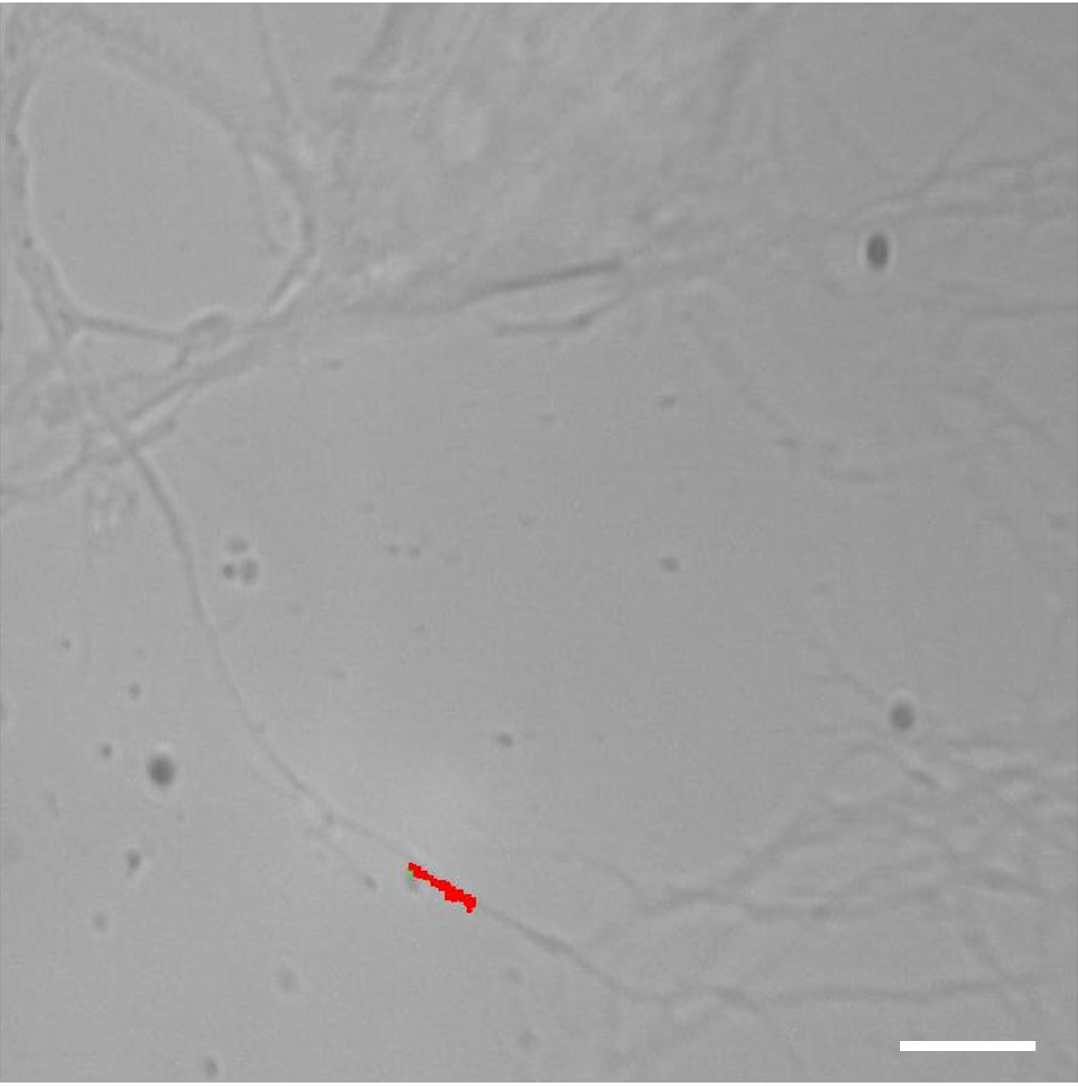} 
   & \includegraphics[width=0.3\linewidth]{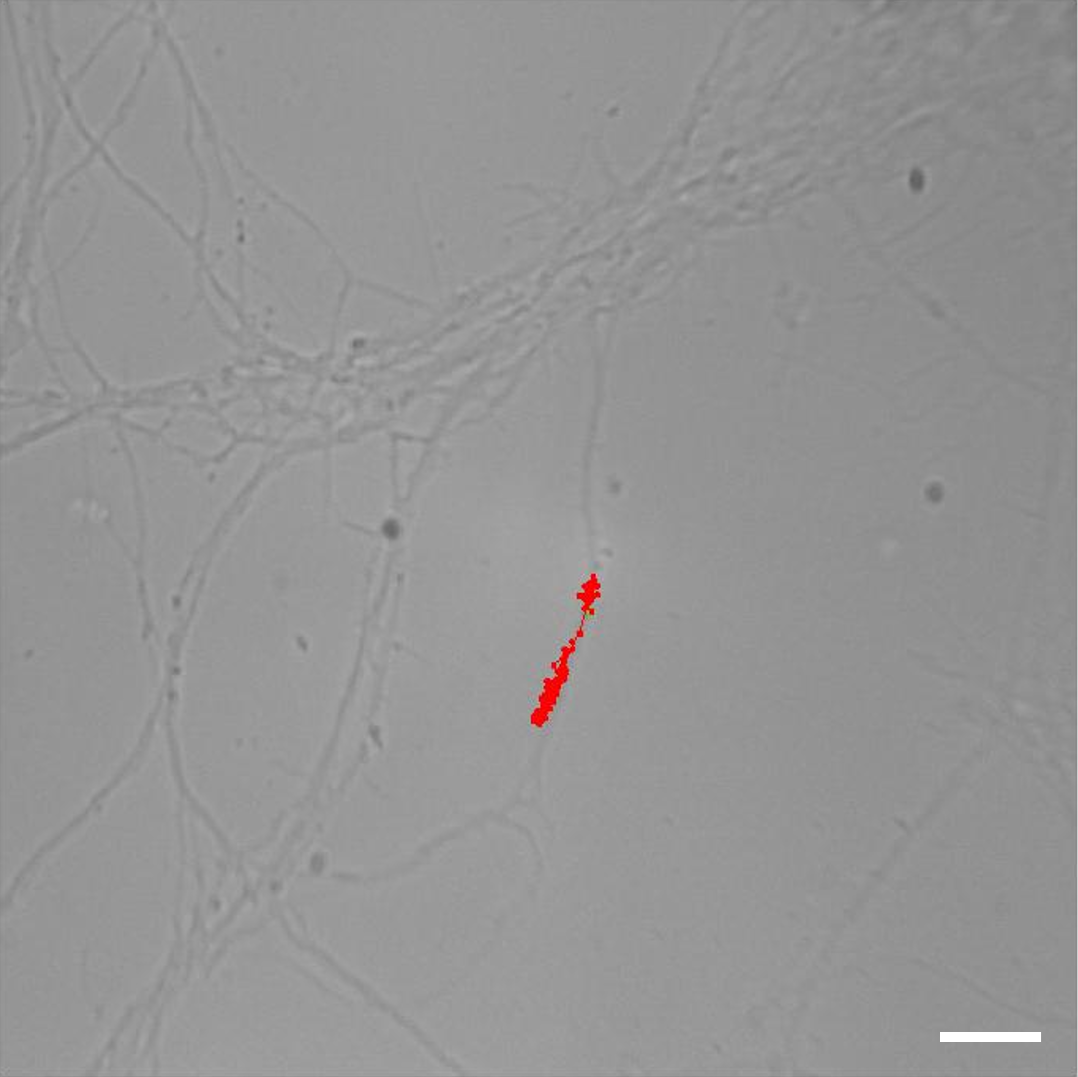} \\

    % regression
    \includegraphics[width=0.33\linewidth, trim=5 0 30 0,clip]{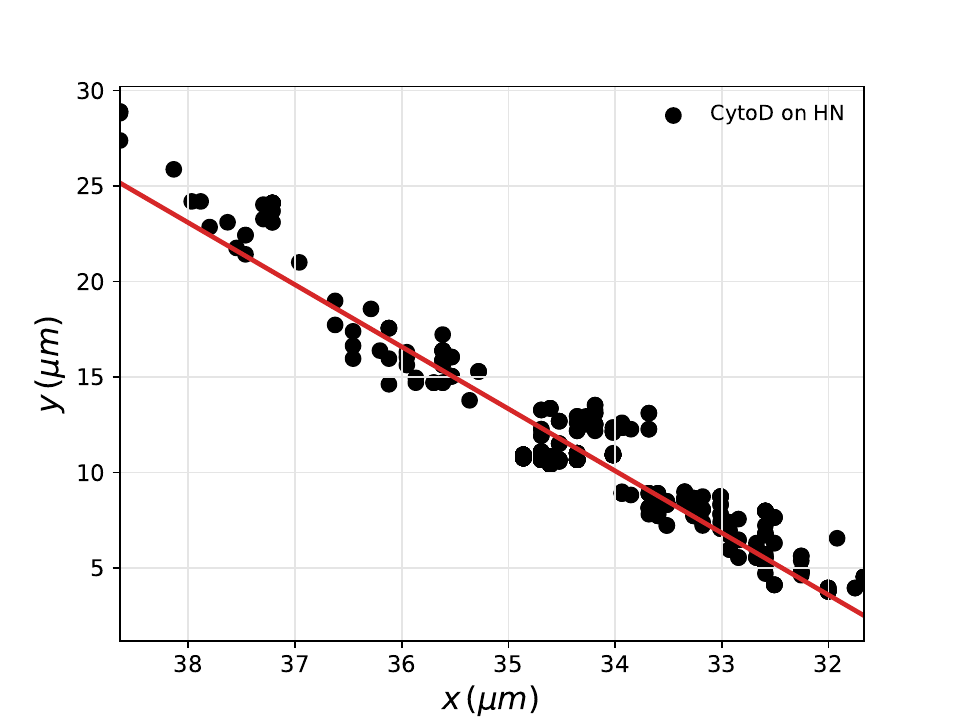} & \includegraphics[width=0.33\linewidth, trim=5 0 30 0,clip]{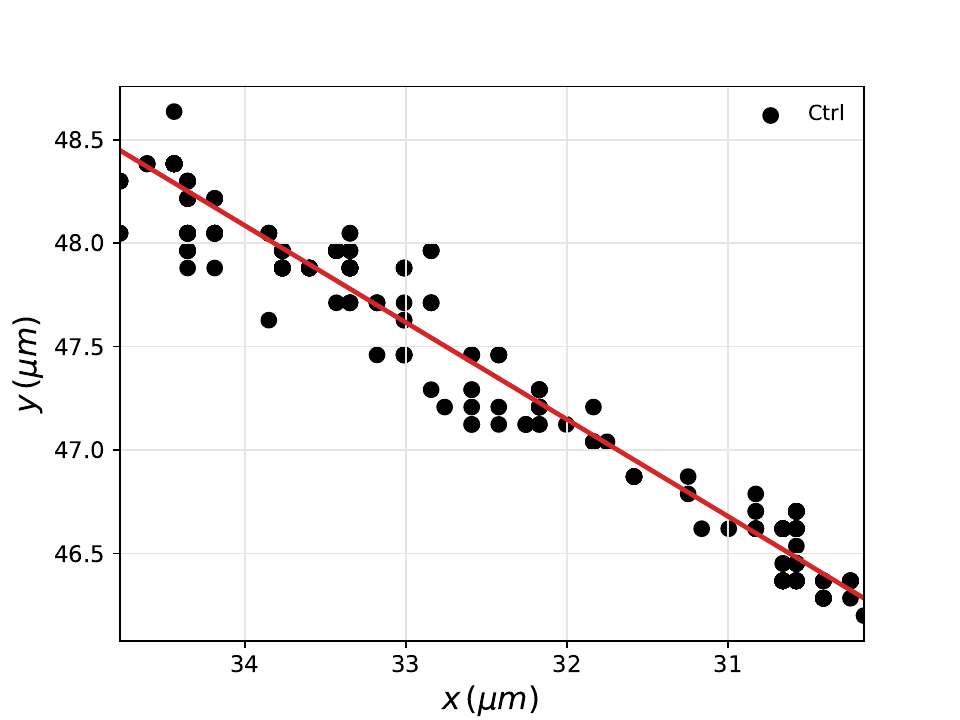}
    & \includegraphics[width=0.33\linewidth, trim=5 0 30 0,clip]{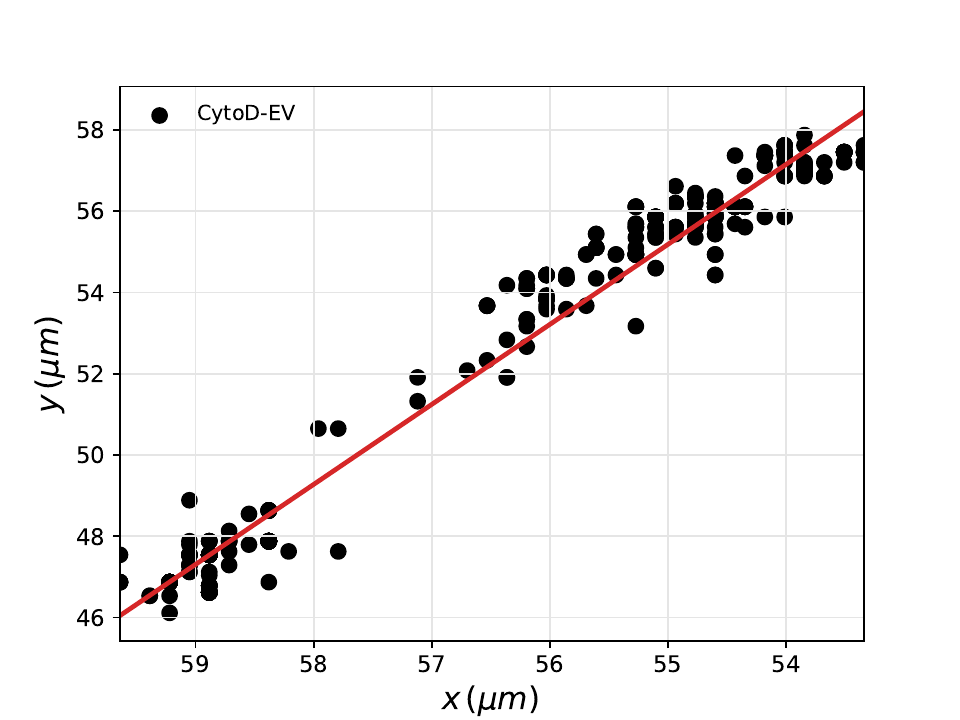}  \\
\end{tabular}
    \caption{Data acquisition and post processing. Top row: example of EVs trajectories (red trace) collected by MATLAB analysis. The EVs are tracked for nearly 4 min. Figures show single frames from EV brightfield recordings of the three datasets: (left) \textit{CytoD-HN}, (center) \textit{Ctrl} and (right) \textit{Cyto-EV} (Scale bar = 10$\mu$m, magnification 63x). Bottom row: The respective main direction of EVs movement has been extrapolated by a linear regression. Experimental data are represented by black dots while the red line is the regression line. Coordinates $(x,y)$ represent the actual position of the vesicle over time, where (0,0) corresponds to the bottom-right corner of the original size image.}
    \label{fig:traj}
\end{figure}

Post-processing the microscopy data, we found that the EVs mostly move tangentially to the neuron surface, so that their motion could be approximated as one-dimensional. To further validate this hypothesis, we compared the mean velocity, computed as the path length over the experiment duration, with the mean velocity obtained by projecting the displacements onto the neuron surface. Thus, to get an estimation of the neurite axis direction, which in good approximation coincides with the main direction of the EV motion, we performed a linear regression on the EV displacements, see Fig. \ref{fig:traj} (bottom row).

The described methods revealed to be almost always effective, except in samples where the EV was almost still and so the  positions remain confined in a very restricted area that the direction found by linear regression was no longer meaningful. In these very few cases, the main direction was retrieved by using an alternative method, consisting of linearly interpolating two extreme points of the EV path.
With the described approach, we found that the component tangent to the neuron surface accounts for nearly $80\%$ of the motion, on average, in all the three datasets, thus supporting the one-dimensional hypothesis. 

As last step before the data analysis, in order to identify directed transport phenomena, we standardized the preferential direction of the EVs towards positive values by inverting the sign of the displacements in samples where the linear regression slope computed on the tangent displacement was negative.

\subsubsection*{Sampling frequency}
The subsampling frequency for the data analysis is chosen by leveraging the autocorrelation function (ACF) analysis of the acquired EVs positions. Indeed, by quantifying the self-similarity of the signal over different time lags, it is possible to verify the suitability of the chosen sampling rate by comparing it with the results obtained from a downsampled signal \cite{vilela2013fluctuation}.
We applied this test to our datasets, comparing the sampling frequency $F = \SI{2}{\hertz}$, i.e. including all video frames, with $F=\SI{0.4}{\hertz}$, i.e. the sampling frequency adopted in \cite{d2021astrocytes} for the data analysis. Since it emerged that the latter was not sufficiently high but at the same time we cannot draw conclusion on the accuracy of the sampling frequency $F = \SI{2}{\hertz}$, we decided to perform the data analysis using all the available video frames as it was the most accurate solution available to us.

\subsection*{Data analysis}
\label{subs::danalysis}
After processing the data with \texttt{MATLAB} as described above, statistical analysis was performed on all the datasets using the \texttt{Python} libraries \texttt{SciPy} and \texttt{Pandas} \cite{van1995python, virtanen2020scipy, mckinney2010data}. Data were first tested for normal distribution and skewness with an accepted level of significance $P \leq 0.05$. Data are expressed as means $\pm$ STD unless otherwise specified. 
We notice that, for what concerns the above mentioned analysis, due to the poor quality of some videos, the dataset has been narrowed to 15 samples for the Ctrl set, 13 for the CytoD-EV set and 14 for the CytoD-HN set.

\subsection*{Mathematical model}
We propose a model to describe the motion of extracellular EVs based on the assumptions contained in D'Arrigo et al. \cite{d2021astrocytes}, according to which vesicle transport occurs through two main mechanisms: \textit{passive} transport, where the vesicle is passively dragged by the cytoskeletal activity, and \textit{active} transport, due to the actin filaments rearrangements inside the vesicle that makes the EVs capable of actively "rolling" over receptors on the neuron surface.
To support the feasibility of the proposed transport mechanism involving PrP interactions with neuronal receptors, in Section \ref{SI::density_estimate} of the Supplementary Information we provide a tentative estimate of the density of proteins per EV and neurite receptors.

\begin{figure}
	\centering
    \includegraphics[width=0.45\textwidth]{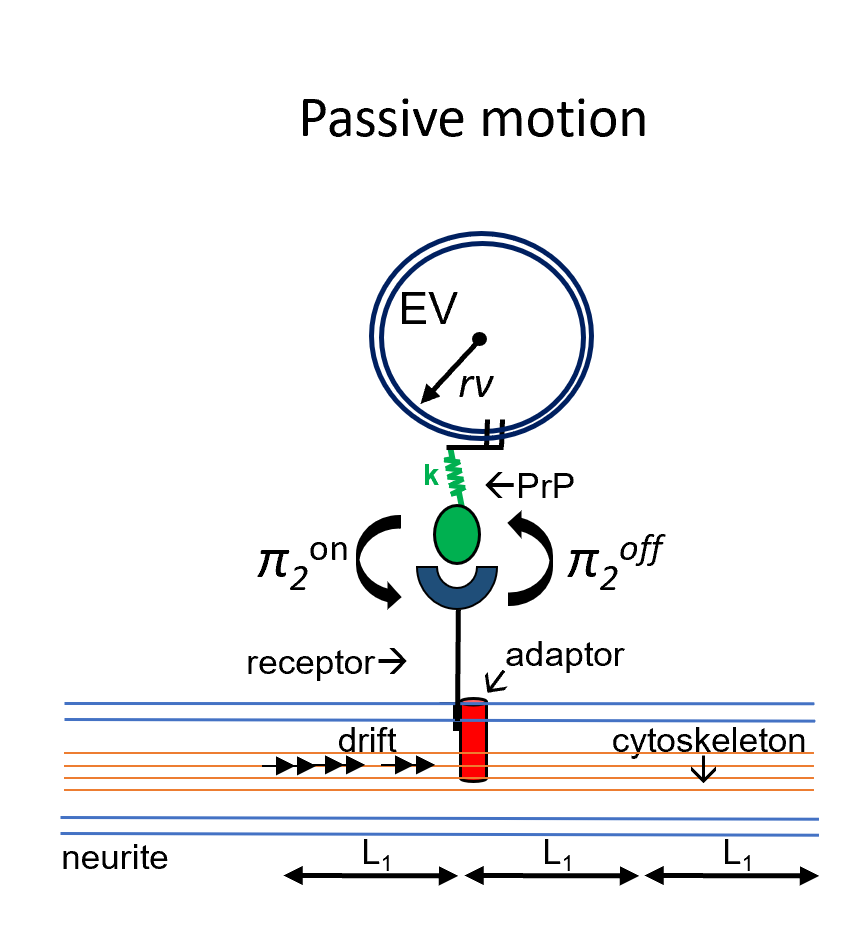}
	\includegraphics[width=0.45\textwidth]{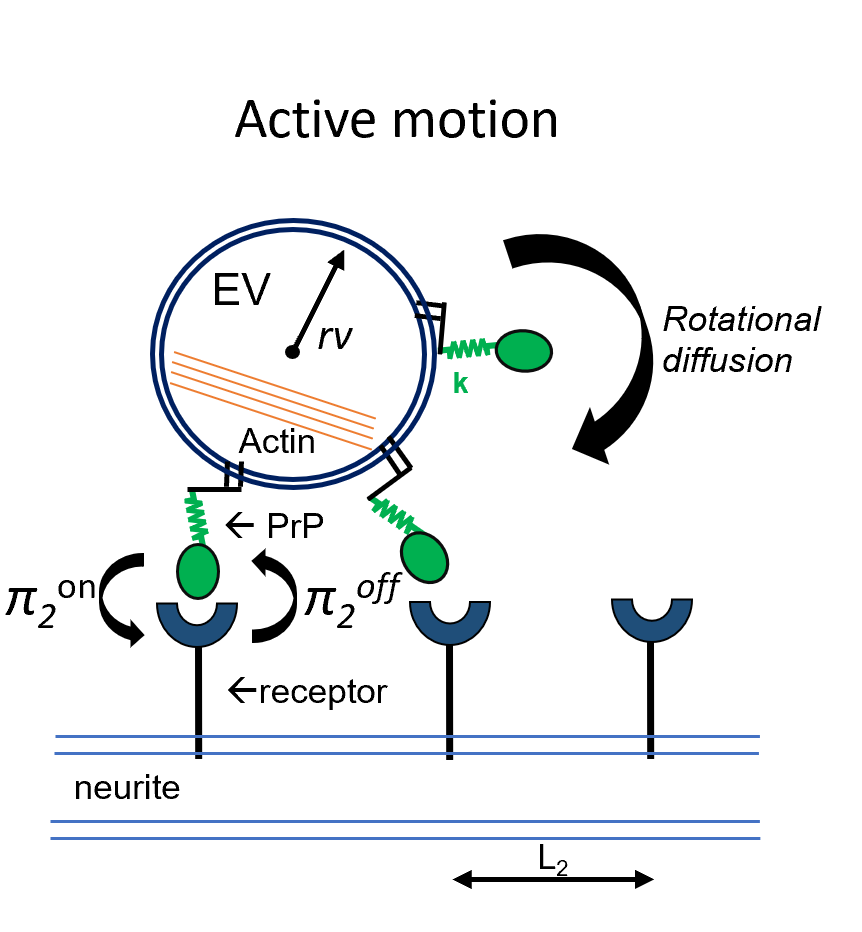}
	\caption{ Schematic representation of the modelled EVs transport mechanisms: (left) passive and (right) active motions.}
	\label{fig:evs_motion}
\end{figure}

The \textit{passive} transport is governed by the behavior of the neuronal receptor to which the EV is linked (see left side of Figure \ref{fig:evs_motion}). In fact, as similarly discussed in \cite{goychuk2014molecular}, the receptor acts as a motor and drags the vesicle (the cargo) along the neurite thanks to a periodic external force field provided by interaction with the polymeric fibers of the cytoskeleton. The cargo is considered to be attached to the motor through an elastic linker that, in our framework, corresponds to the prion protein. The EV and the receptor are coupled by an elastic force, $f_{el} = k(y-x)$, where $x$ and $y$ are, respectively, the coordinates of the EV and the receptor along the neurite, and $k$ is a spring constant, whose value is assumed constant and should depend on the average PrP conformations, i.e. extended brush length and persistence length  \cite{marbach2022nanocaterpillar}. The receptor also undergoes diffusion along the neuron membrane and we assume that its diffusion coefficient, $D_r$, is related to its viscous friction coefficient $\xi_r$ by Einstein relation, $D_r=k_BT/\xi_r$ where $k_B$ is the Boltzmann constant and $T$ is the absolute temperature. Likewise, also the vesicle diffuses in the extracellular bath (which, to a good approximation, has the same physical features of water) with diffusion coefficient $D_v=k_BT/\xi_v$, where $\xi_v$ is a viscous friction coefficient. For simplicity, we consider both the vesicle and the receptor to be spherical with radius $r_v$ e $r_r$, respectively, so that their friction constants are related to their radius by Stokes' law $\xi_{v}=2\cdot6\pi\eta_{w} r_{v}$, $\xi_{r}=2\cdot6\pi\eta_{c} r_{r}$ where $\eta_c$ and $\eta_w$ are the viscosity coefficients of cytosol and water \cite{marbach2022nanocaterpillar,goychuk2014molecular}. Finally, the stochastic nature of the receptor motion is taken into account with a flashing ratchet model, i.e, we assume that the receptor is constantly transitioning between two potentials: a one-dimensional, periodic, asymmetric saw-tooth function $V_1$ of period $L_1$ that accounts for a directed motion, and a vanishing  potential $V_0 = 0$ that allows the receptor to diffuse freely. Then, the system evolves according to a set of coupled stochastic equations so that, in general, a biased diffusion ensues
\begin{subequations}
\begin{align}
\xi_v \,dx & = -k(x-y)\,dt + \sqrt{k_BT\xi_v} \,dW_v  \label{eq:EV_passive}, \\[2mm]
\xi_r \,dy& = -S_1(t) V_1'(y)\,dt + \sqrt{k_B T\xi_r} \, dW_r  \label{eq:receptor_passive},
\end{align}
\label{eq:passive}
\end{subequations}
where $W_{v}$, $W_{r}$ are two independent Brownian processes. In Eq.(\ref{eq:receptor_passive}), $S_1(t)$ is a continuous-time Markov process that represents the two possible states of the neuronal receptor: $S_1=1$ if the receptor is bound to the cytoskeleton, while $S_1=0$ if the receptor is detached from the cytoskeleton. Therefore, when $S(t)=1$, the receptor is under the influence of the potential $V_1(y)$, which we explicitly write as
\begin{equation}
    V_1(y)=\begin{cases}
        h_1\dfrac{y}{\alpha}&\quad\text{if   \quad }0\leq y\leq \alpha L_1, \\[10pt]
        h_1\dfrac{L_1-y}{1-\alpha}&\quad\text{if  \quad }\alpha L_1< y\leq L_1,
    \end{cases}
\end{equation}
where $h_1$ is the peak value of the potential representing the height of the energy barrier that the receptor must overcome to slide along the F-actin cytoskeletal network, $L_1$ is its period, and $\alpha$ is a shape parameter in $(0,1)$ that measures the asymmetry of the potential. When $\alpha=1/2$, the potential is symmetric, a condition that we generally avoid in this case, so that we limit our analysis to $0 < \alpha < 1/2$. It is worth to notice that the elastic force $f_{el}$ does not appear in Eq.(\ref{eq:receptor_passive}) because it is assumed to be balanced by the strong binding reactive forces exerted by the cytoskeleton. 

The \textit{active} transport of the EVs is due to its capability to change shape through actin filaments rearrangements, so that it is able to step from one receptor to one nearby.  For the sake of simplicity, we assume the receptors to be uniformly distributed along the neurite, with a spacing of approximately the diameter of the EV, so that they are easily reachable by the PrP legs of the vesicle (see right side of Figure \ref{fig:evs_motion}). The particle is assumed to be always bound to one of the neural receptors, hence, as in the passive transport, it is subject to an elastic force $f_{el} = k(y-x)$. The stepping from a receptor to a neighboring one has no preferential direction, so we decided to model the receptor motion with a flashing Brownian ratchet that switches between the vanishing potential $V_0$ and a \textit{symmetric} saw-tooth potential $V_2$, in contrast with the \textit{asymmetric} one used for the passive case. We introduce a new variable $y_\text{act}$, that represents the position of the active receptor, i.e., the receptor that is currently linked to the vesicle. Similarly to the previous case, we posit the following equations for the EV position $x$, and the active receptor $y_\text{act}$
\begin{subequations}
\begin{align}
\xi_v\,dx &= -k(x-y_\text{act})\,dt + \sqrt{k_BT\xi_v}\,dW_v, \label{eq:EV_active} \\[2mm]
dy_\text{act} & = -S_2(t)\frac{V_2'(y_\text{act})}{\xi_\text{eff}}dt + \sqrt{D_r}\,dW_r, \label{eq:receptor_active}
\end{align}
\label{eq:active}
\end{subequations}
where $D_r$ is a rotational diffusion coefficient that we take to be equal to $\frac{k_BT}{\xi_\text{eff}}$, with $\xi_\text{eff}$ computed as the average between $\xi_v$ and $\xi_r$.  Unlike Eq.\eqref{eq:receptor_passive}, and despite its appearance, Eq.\eqref{eq:receptor_active} is not a momentum balance equation for the overdamped receptor motion. It simply states that the position of the receptor currently linked to the vesicle can change in a small time step $dt$ for two reasons: a rotational diffusion of the vesicle and an elongation of the vesicle shape through its internal actin filament rearrangements, whose effect is modeled with a symmetric sawthooth potential, analogously with the passive case. The constants $k$ and $\xi_v$ are the same introduced for the passive transport, $W_v$ is a Brownian process and $S_2(t)$ represents a Markov process with values 1 and 0 that models the random binding and unbinding process. Specifically, the symmetric periodic potential $V_2$ is taken to be
\begin{equation}
    V_2(y_\text{act})=\begin{cases}
       2 h_2\, y_\text{act}&\quad \text{ if  }\quad 0\leq y_\text{act}\leq L_2/2 \\[2mm]
       2 h_2 \,(L_2-y_\text{act})&\quad \text{ if  }\quad L_2/2\leq y_\text{act}\leq L_2
    \end{cases}
\end{equation}
where $h_2$ is the peak value, representing the height of the energy barrier that the vesicle must overcome to break the bond with the receptor to which it is currently linked, and $L_2$ is the period of the potential, approximately equal to the distance between the receptors.
\begin{figure}
\begin{center}
\includegraphics[width=0.6\linewidth]{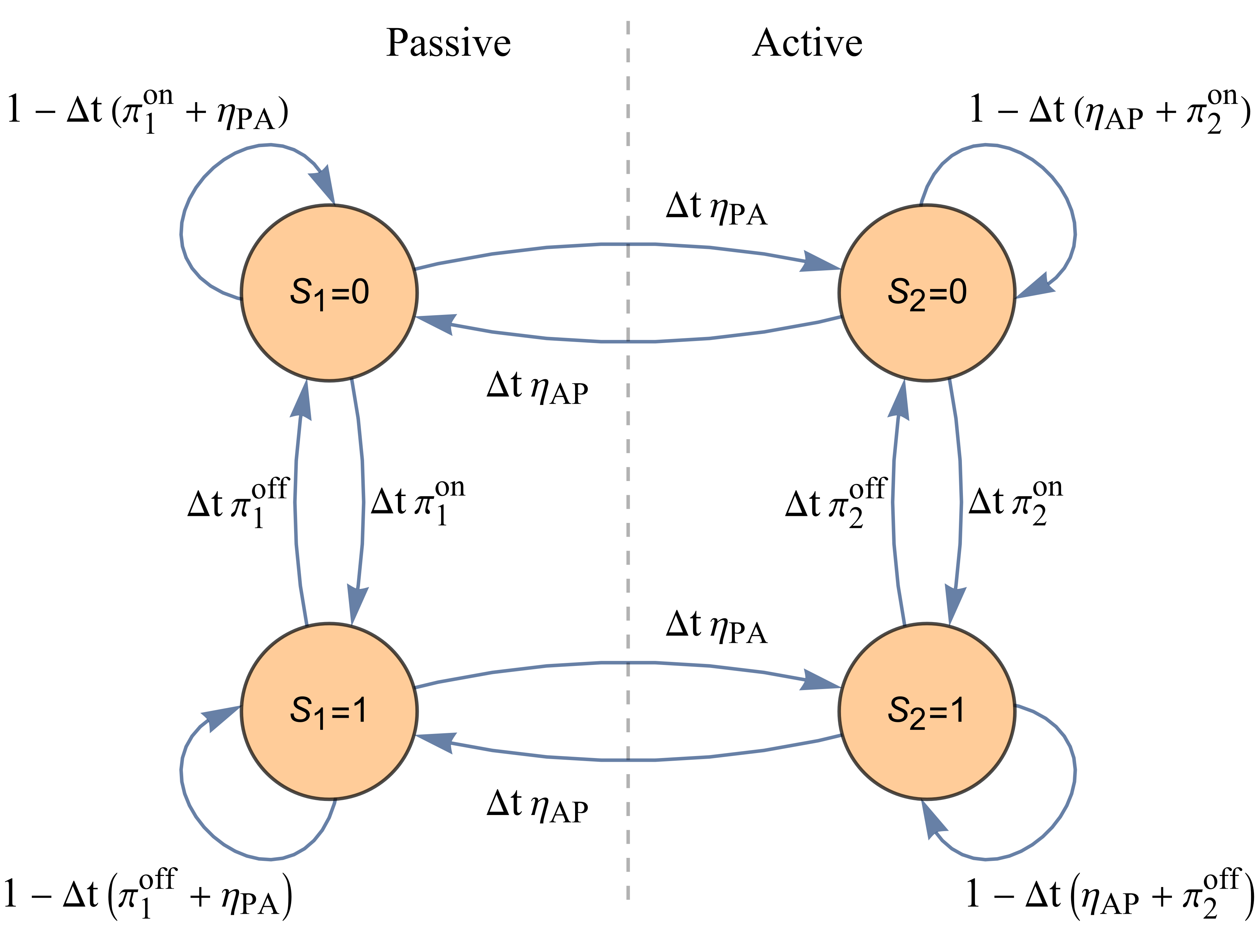}
\end{center}
\caption{Graph describing the transition probabilities in a small time-step $\Delta t$ of the overall Markov process comprising $S_1(t)$, $S_2(t)$ and the switching between transport models. The binding and unbinding rates in the passive model are $\pi_1^\text{on}$ and $\pi_1^\text{off}$, while in the active model are $\pi_2^\text{on}$ and $\pi_2^\text{off}$. The two models are stochastically coupled with switching rates $\eta_{PA}$ and $\eta_{AP}$.}
\label{fig:MCgraph}
\end{figure}

The overall motion of EVs is a combination of the passive and active mechanisms described above. The binding and unbinding probabilities are obtained as the product of the binding and unbinding rates $\pi^{\text{on}}_k$ and $\pi^{\text{off}}_k$ ($k=1,2$ for, respectively, passive and active models) with a small time-step $\Delta t$, the corresponding Markov chain is described in Fig.\ref{fig:MCgraph}. To account for the possible random switching among passive and active transport, we also consider the transition rates $\eta_{PA}$ (from passive to active) and $\eta_{AP}$ (from active to passive).

\subsection*{Numerical method}
The experimental datasets include three different scenarios: (1) control (Ctrl), (2) neurons treated with CytoD (CytoD-HN) and (3) EVs treated with CytoD (CytoD-EV). In the mathematical model, these scenarios are distinguished by the presence of different vesicle transport mechanisms, namely: (1) both passive and active transport, (2) only active transport and (3) only passive transport (see Fig.\ref{fig:MCgraph}). In case (1) the system can be in any of the four states of Fig.\ref{fig:MCgraph}, in case (2) only transitions between states $S_2=0$ and $S_2=1$ can occur; in case (3) only transitions between states $S_1=0$ and $S_1=1$ can occur. 

The complete algorithm for case (1), when both active and passive transport are present, is described below.
\begin{enumerate}
\item  {[\textbf{initialization}]} At $t=0$, we choose one of the possible states $S_1=0$, $S_1=1$, $S_2=0$ or $S_2=1$ with equal probability and initialize the position of the particle and the receptor as $x_0=0$ and $y_0=0$ (or $y_{act0} = 0$), respectively. 
\item {[\textbf{displacement step}]} To a given state, $S_k = 0;1$, it corresponds a system of equations (either \eqref{eq:passive} or \eqref{eq:active}) that give the displacements of the vesicle and the receptor in a time-step $\Delta t$. To update the values of $x$ and $y$ (or $y_{act}$), we use a standard Euler-Maruyama scheme \cite{toral2014,bressloff2014}.  Normally distributed random numbers with expected value zero and variance $\Delta t$ are used for $dW_r$ and $dW_v$. 
\item {[\textbf{transition step}]} Given the state $X_t$ of the chain at time $t$, the next state of the Markov chain is determined according to the transition probabilities of Fig. \ref{fig:MCgraph}. More precisely, we partition the interval $[0,1]$ into sub-intervals of length proportional to the transition probabilities from the current state and draw a uniformly distributed random number $R$ in $[0,1]$. The state at time $t+\Delta t$ is determined by finding which sub-interval contains $R$. For example, starting from $X_t=$``$S_1=0$'', the next state will be determined as follows
\begin{equation}
X_{t+\Delta t} = 
\begin{cases}
S_1=1 \qquad & \text{ if } \quad R \in \big[0, \pi_{1}^{\text{on}}\Delta t \big), \\[2mm]
S_2=0 \qquad & \text{ if } \quad R \in \big[\pi_{1}^{\text{on}}\Delta t , (\pi_{1}^{\text{on}} + \eta_{PA})\Delta t \big), \\[2mm]
S_1=0 \qquad & \text{ if } \quad R \in \big[(\pi_{1}^{\text{on}} + \eta_{PA})\Delta t, 1\big].
\end{cases}
\end{equation}
\item Repeat from step 2, until we reach the desired number of steps.
\end{enumerate}

In order to simulate the two CytoD treated sets (cases (2) and (3) above), it is sufficient to set both $\eta_{PA}$ and $\eta_{AP}$ to zero and to choose the correct initial state. In this way, one of the two transport mechanisms will be excluded from the whole simulation.

Finally, we remark that, since the systems of stochastic equations (\ref{eq:passive}, \ref{eq:active}) includes different characteristic times, related to the transition rates of the  Markov processes, the simulation time-step $\Delta t$ must obviously satisfy the following constraint
\begin{equation*}
    \Delta t<\min (1/\pi_{1}^\text{on},1/\pi_{1}^\text{off},1/\pi_{2}^\text{on},1/\pi_{2}^\text{off},1/\eta_\text{AP},1/\eta_\text{PA}).
\end{equation*}

\section{Results}
\label{sec::results}
In this Section we provide the results both from the data analysis on the experimental outcomes and from the numerical simulation of the mathematical model proposed in Section \ref{sec::methods}. 
The data used in this study and the codes for EVs tracking and numerical simulations are available on \texttt{Zenodo}:
\href{https://zenodo.org/records/15388642}{zenodo-evs}.

\subsection*{Experimental Results}
For all the experimental results reported in the present Section, due to image quality reasons, the dataset has been narrowed to 15 samples for the Ctrl set, 13 for
the CytoD-EV set and 14 for the CytoD-HN set.

\subsubsection*{Normality and Skewness tests}
From a qualitative analysis on the video frames, the vast majority of EVs trajectories did show neither marked trends nor smooth behaviors but rather high frequency oscillations. This lead us to suppose their motion to be characterized by a strong stochastic component. We therefore tested the hypothesis that their motion was completely Brownian by performing a two-sided Kolmogorov-Smirnov test, with null hypothesis that the EVs point-to-point displacements were drawn from a Normal distribution. To this purpose, we removed zero-length point-to-point displacements from each of the three dataset. With a confidence level of 95$\%$, the displacements distributions cannot be considered Gaussian.
To further elucidate the characteristics of the motion, we performed a skewness test to investigate whether there is a drift component in the motion of EVs in any of the datasets. 
The analysis shown that both the Ctrl (\textit{p-value} = 0.00296, $n = 15$) and CytoD-EV (\textit{p-value} = 0.027501, $n = 13$) present a directed transport component in the motion, while the set CytoD-HN (\textit{p-value} = 0.73127, $n = 14$) has no drifting component.

\subsubsection*{Displacement and velocity analysis}
To our purposes, we introduce a set of kinematic indicators, reported in Table \ref{tab::indicators}, to quantify the EVs motion. 
\begin{table}
	\centering
\begin{tabular}{lp{11cm}}
\toprule
\textbf{Indicator} & \textbf{Definition} \\
\midrule
path-length & the sum of the absolute value of the displacements over the whole time of the experiment\\
\hline
mean velocity & the ratio between the path length and the time length of the experiment \\
\hline
MF mean velocity & the ratio between the path length and the sum of the time steps corresponding to a non-zero displacement (i.e. the moving frames, MF) \\
\hline
net mean velocity & the mean velocity obtained considering the sign of the displacements instead of their absolute value\\
\hline
zero velocity rate & the ratio of frames with zero displacement to the total number of frames\\
\hline
mean displacement & the average of the absolute value of the displacements\\
\hline
MF mean displacement & the mean displacement computed considering only the moving frames \\
\hline
explored distance & the maximum traveled distance defined as \newline $\max_{t\in[0,T]}x(t)-\min_{t\in[0,T]}x(t)$, \newline
where $x(t)$ denotes the spatial coordinate of the particle at time $t$\\
\bottomrule
\end{tabular}
	\caption{Set of kinematic indicators introduced to quantitatively describe the EV motion. All the quantities refer to the component of the motion tangent to the neuron surface.}
 \label{tab::indicators}
\end{table}
The results of the statistical analysis are then summarized in Table \ref{tab::analysis_indicator}.

\begin{table}
\centering 
    \begin{tabular}{l l c c c c}
    
    \toprule
    \multicolumn{2}{c}{ \textbf{Indicator} }
    & \multicolumn{3}{c}{ \textbf{Data sets} } & \textbf{Kruskal test}\\
    
     Name & UM & Ctrl & CytoD-EV & CytoD-HN & Significance\\
    
    \midrule
pathlength & $\mu$m & 188.7$\pm$163.2 & 399.6$\pm$284.9 & 295.6$\pm$214.9 &   \small{* (AB)}
\\
mean velocity & $\mu$ms$^{-1}$ & 0.150$\pm$0.129 & 0.357$\pm$0.268 & 0.295$\pm$0.229 & \small{* (AB, AC)} 
\\
MF mean velocity & $\mu$ms$^{-1}$ & 0.762$\pm$0.350 & 0.705$\pm$0.350 & 0.775$\pm$0.195 & - 
\\
net mean velocity & $\mu$m$^{-1}$ & 0.005$\pm$0.004 & 0.006$\pm$0.005 & 0.006$\pm$0.004 & - 
\\
zero velocity rate & $ - $ & 0.82$\pm$0.12 & 0.54$\pm$0.27 & 0.65$\pm$0.22 & \small{** (AB); * (AC)}  \\
mean displacement & $\mu$m & 0.075$\pm$0.064& 0.178$\pm$ 0.134 & 0.148$\pm$0.115 & \small{* (AB, AC)}
\\
MF mean displacement & $\mu$m & 0.1$\pm$0.094& 0.565$\pm$0.679 & 0.338$\pm$0.422 & \small{** (AB); * (AC)} 
\\
explored distance & $\mu$m & 11.49$\pm$6.9 & 13.60$\pm$7.2 & 12.2$\pm$4.5 & - \\
\bottomrule
\end{tabular}
\caption{Indicators values (mean$\pm$stdev) computed for each experimental conditions in the dataset. In the last column of the table the results of the Kruskal statistical test are reported. Symbols (*) and (**) denote cases where the \textit{p-value (p)} of the test is $p<0.05$ and $p<0.01$, respectively. The dataset has been consist of 15 samples for the Ctrl set, 13 for
the CytoD-EV set and 14 for the CytoD-HN set. (AB) indicates that the level of significance is referred to the comparison between the Ctrl and the CytoD-EV population, while (AC) for comparison between the Ctrl and the CytoD-HN population.}
\label{tab::analysis_indicator}
\end{table}

\begin{figure}
    \centering
    \includegraphics[width=0.33\linewidth,trim= 0 0 25 10,clip]{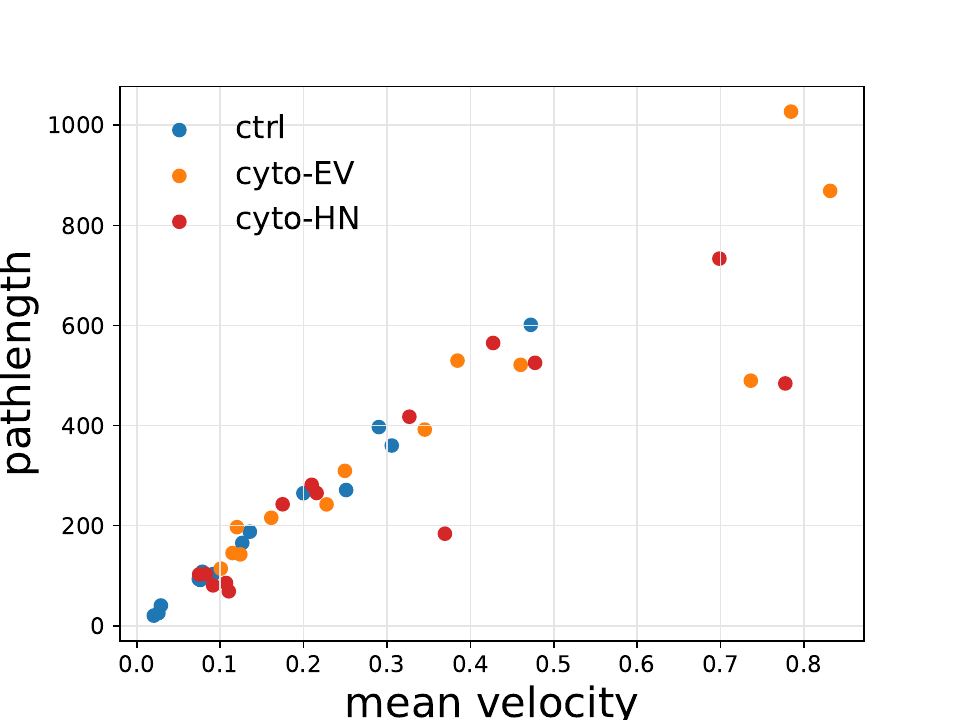}\hfill
    \includegraphics[width=0.33\linewidth,trim= 0 0 25 10,clip]{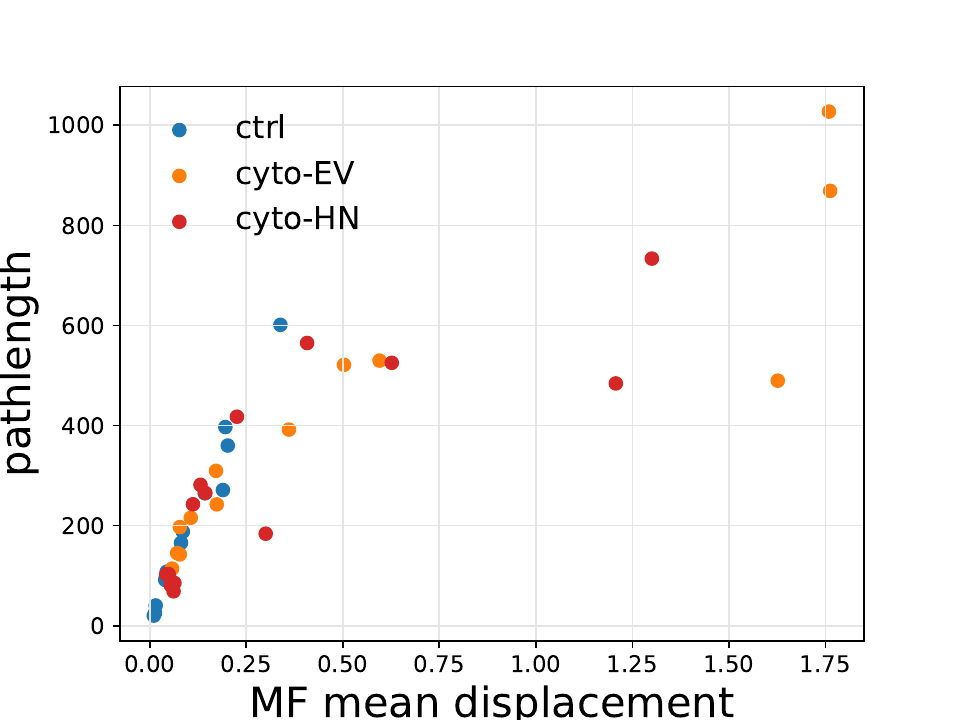}\hfill
    \includegraphics[width=0.33\linewidth,trim= 0 0 25 10,clip]{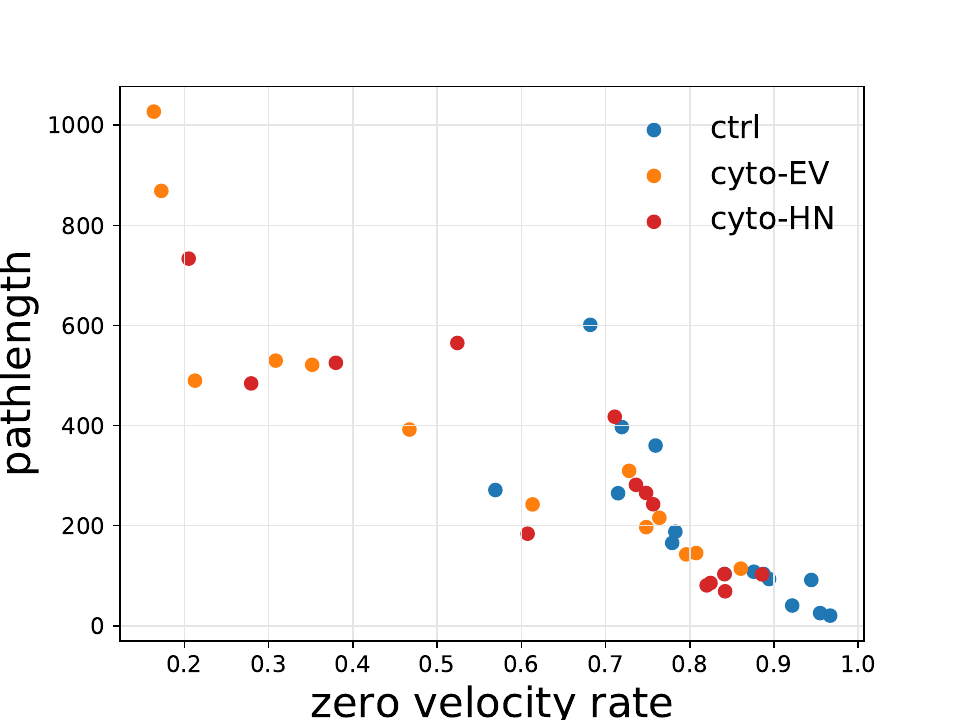}\hfill
    \includegraphics[width=0.33\linewidth,trim= 0 0 25 10,clip]{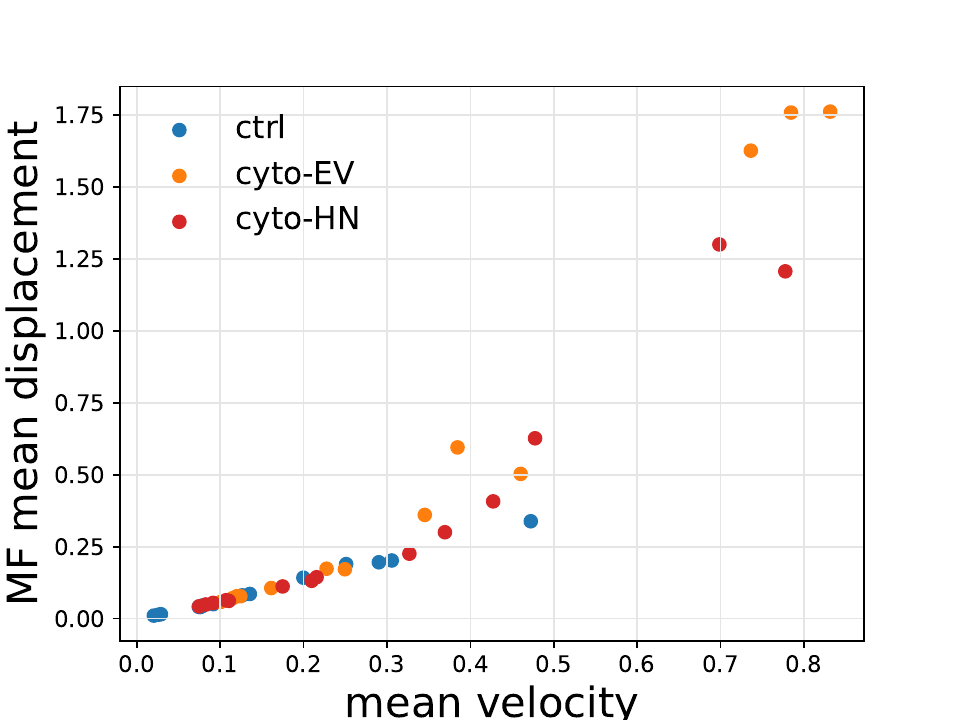}\hfill
    \includegraphics[width=0.33\linewidth,trim= 0 0 25  10,clip]{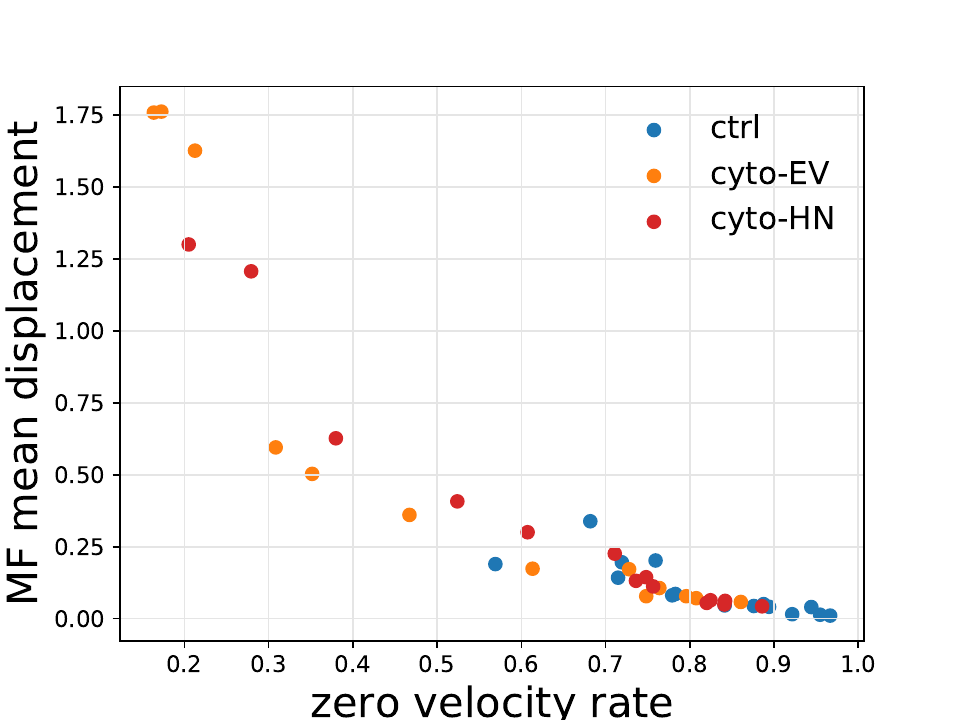}\hfill
    \includegraphics[width=0.33\linewidth,trim= 0 0 25  10,clip]{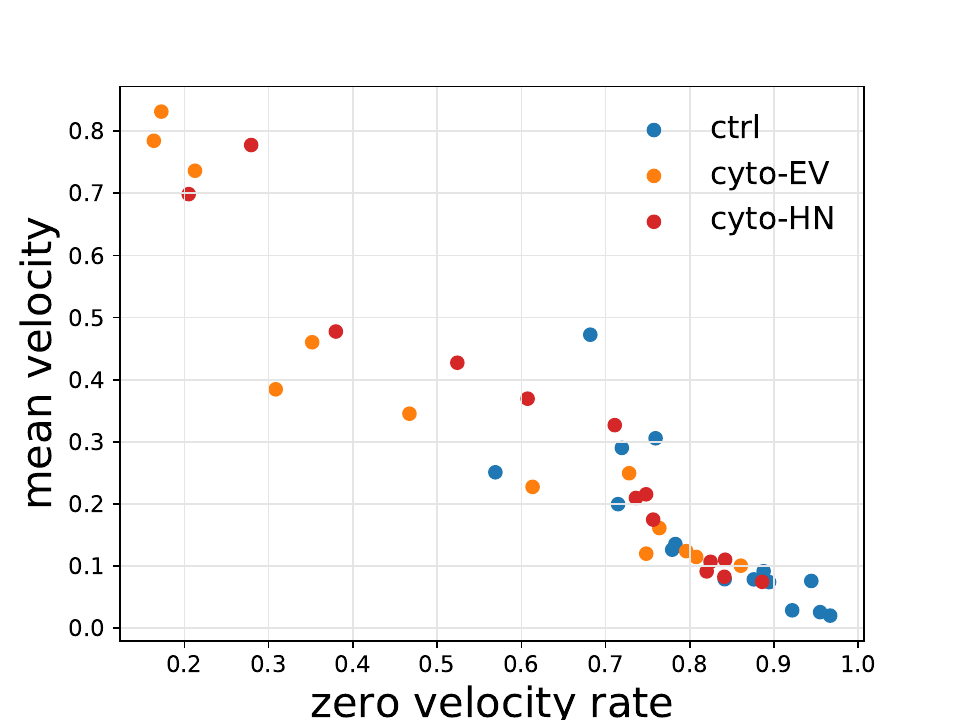}\hfill
    \caption{Scatter plot illustrating the relation between couples of kinematic indicators for the three dataset Ctrl (blue), CytoEV (orange) and CytoHN(red). A subset of the possible indicator combinations has been selected according to the level of significance of the Kruskal test in Table \ref{tab::analysis_indicator}}
    \label{fig:scatter}
\end{figure}

We observe that, overall, basing on both the indicator values and the outcomes of the Kruskal statistical test on the population median values, Ctrl is the set that differs significantly from the others.
In particular, the Ctrl set has a significantly higher \textit{zero velocity rate} (\textit{p-value} = 0.0046 for the \textit{Ctrl}/\textit{Cyto-EV} test; \textit{p-value} = 0.0234 for the \textit{Ctrl}/\textit{Cyto-HN} test) compared to the two populations treated with Cytochalasin D. This trend obviously reflected in a lower \textit{mean velocity} (\textit{p-value} = 0.0137 for the \textit{Ctrl}/\textit{Cyto-EV} test; \textit{p-value} = 0.0362
 for the \textit{Ctrl}/\textit{Cyto-HN} test) and \textit{pathlength} (\textit{p-value} = 0.013720613
 for the \textit{Ctrl}/\textit{Cyto-EV} test) for the Ctrl set.
A further support to the evident reduced motility that characterizes the Ctrl set with respect to the CytoEV and CytoHN sets, we notice a lower \textit{MF mean displacement} and \textit{mean displacement}.
On contrary, no distinctions among the three population is possible on the basis of the \textit{MF mean velocity}, \textit{net mean velocity} and \textit{explored distance} indicators.
Consistently with the outcomes of the Kruskal test in Table \ref{tab::analysis_indicator}, a greater characterization of the non treated samples compared to the ones treated with Cytochalasin D emerges also from the scatterplots in Fig. \ref{fig:scatter}, where the clustering of the Ctrl set is evident.
Moreover, we notice a non trivial weak negative correlation between the \textit{zero velocity rate} and the \textit{MF mean displacement} indexes, see Fig. \ref{fig:scatter} (bottom-center). Conversely, we do not observe a clear negative correlation between the \textit{zero velocity rate} and \textit{pathlength} indexes as may be expected, see Fig. \ref{fig:scatter} (top-right).

\begin{figure}
    \centering
    {\includegraphics[width=0.53\textwidth, trim= 10 0 25  10,clip]{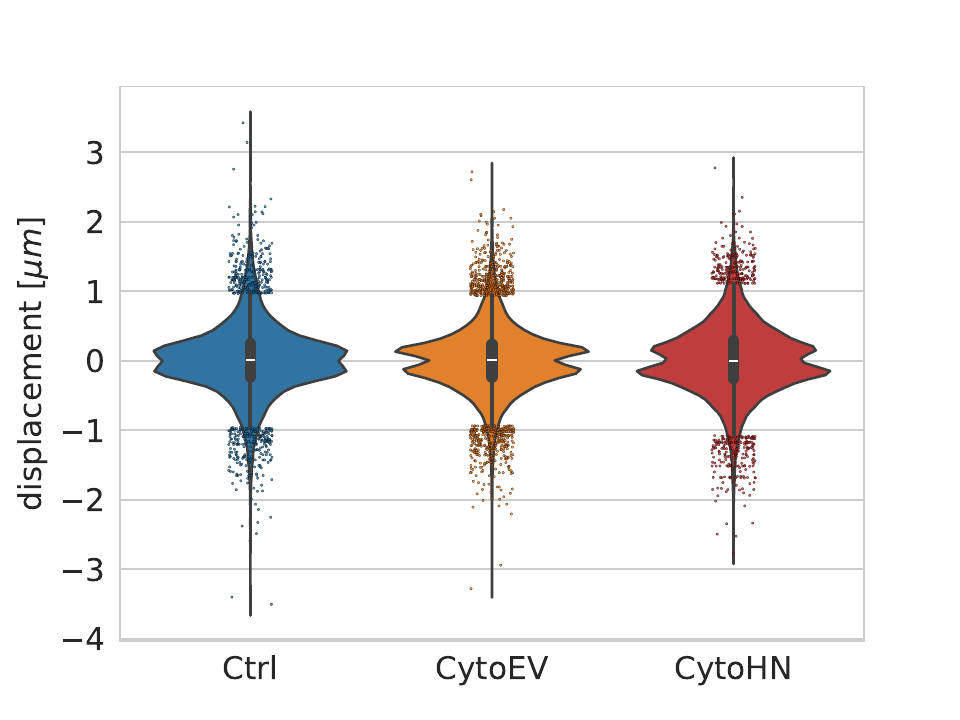}
    \includegraphics[width=0.46\textwidth, trim= 10 0 25  10,clip]{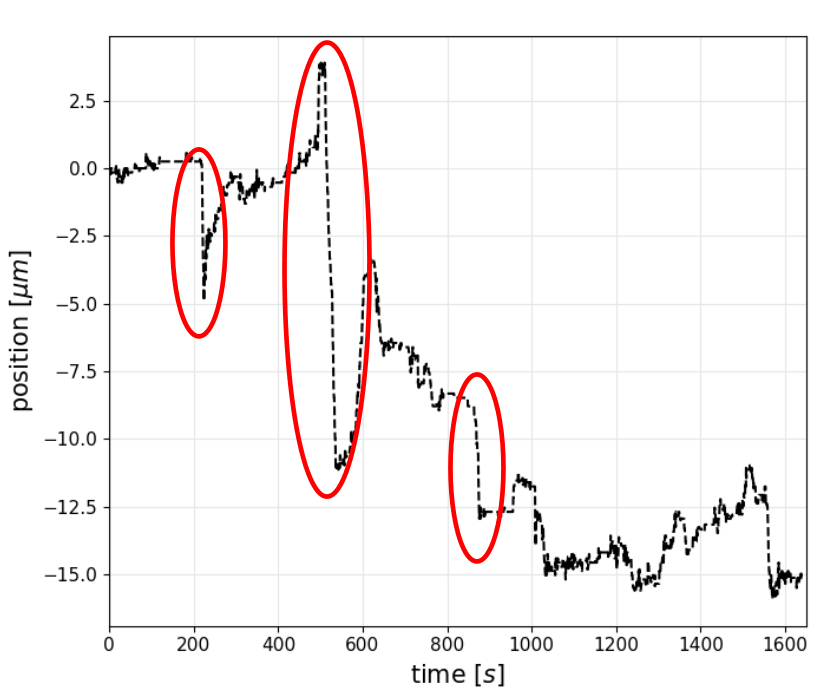}
    }
    \caption{(Left) Violin plot of the displacements absolute value for each dataset: Ctrl (blue), CytoD-EV (Orange) and CytoD-HN (red). Superimposed dots represents the outliers of each population. (Right) Example plot of the position over time for an EV selected from the CytoD-EV set. Red ovals highlight abrupt jumps in the EVs position, supposedly driven by cytoskeletal transport phenomena.}
    \label{fig:large_steps}
\end{figure}

The possible existence of a drift term in the Ctrl and CytoD-EV was also investigated.
Indeed, at a first sight, in several cases the Ctrl and the CytoD-EV trajectories evidence that the directed transport is characterized by the presence of sudden large movements that make the EVs travel a relatively long distance in a matter of seconds, as shown in Fig. \ref{fig:large_steps} (right).
Although quite rare, these events have a strong impact on the overall directionality of movement. As so, it can be reasonably assumed that these rapid shifts most likely belongs to the tails of the non-zero displacements distribution in Fig. \ref{fig:large_steps}  (left). Specifically, rapid long movements shown in Fig. \ref{fig:large_steps}(right) are the distributions outliers, which, although present in each of the  three populations, are significantly more prominent in the Ctrl ($6,5\%$) and in the CytoD-EV ($4,95\%$) sets than in the CytoD-HN population ($3,9\%$).

To conclude, although the proposed data analysis is not conclusive due to population variability and limited populations size, it does show interesting trends. In fact, the Ctrl and CytoD-EV sets appear to have a stronger drift component compared to CytoD-HN, suggesting potential differences in their underlying physical mechanisms of movement.
This observation further supports the hypothesis put forward by D'Arrigo et al. in \cite{d2021astrocytes}, which is at the basis of the proposed mathematical model, according to which the cytoskeletal activity is involved in the transport mechanism of the EVs, playing a major role in imposing a drift.

\subsection*{Numerical Results}
\subsubsection*{Parameters settings}

The model parameters, whenever available, are estimated from available experimental data either by direct measurements or by well-established physical relations, e.g. in the case of the Stokes friction coefficients $\xi_v$ and $\xi_r$, and are collected in Table \ref{table:physical_parameters}.
On the other hand, other parameters of the model, such as those that determine the shape of the ratchet potentials or the transition rates of the Markov processes, cannot be directly related to experimentally measured physical quantities and are therefore at best calibrated to match the experimental data, see Table \ref{table:fitted_parameters}. We remark that the values indicated in Table \ref{table:fitted_parameters} for the $\eta_{AP}$ and $\eta_{PA}$ should be limited to the simulation of the Ctrl condition. Both the treatment cases, CytoD-EV and CytoD-HN, are indeed reproduced by setting $\eta_{AP} = \eta_{PA} = 0$.

\begin{table}
\centering 
    \begin{tabular}{l l l p{8cm}}
    \toprule
     \textbf{Parameter} &  \textbf{UM} & \textbf{Value} & \textbf{Method and references}
\\
 \midrule
 
$k_B$  & J/K & $1.380649\cdot10^{-23}$ & \cite{gupta2020units}
\vspace{0.1cm} \\
$T$  & K & 310.15 & experiments
\vspace{0.1cm} \\
$\eta_w$ & kg/(ms) & $6.913\cdot10^{-4}$ & \cite{korson1969viscosity}
\vspace{0.1cm} \\
$\eta_c$ & kg/(ms) & $6.9\cdot10^{-3}$ & (rough estimate) x$\left[10 - 1500\right] \eta_w$  depending on the protein size \cite{goychuk2014molecular}
\vspace{0.1cm} \\
$r_v$  & m  & $300\cdot10^{-9}$ & experiments 
\vspace{0.1cm} \\
$r_r$ & m & $7\cdot10^{-9}$ &  (rough estimate) \cite{riesner1996disruption}
\vspace{0.1cm} \\
$\xi_v$ & N$\cdot$s/m & $7.8184\cdot10^{-9}$ & hydrodynamic friction of spherical particles in viscous fluids \cite{stokes1851effect}: $2\cdot6\pi\eta_w r_v$
\vspace{0.1cm}\\
$\xi_r$ & N$\cdot$s/m & $1.8209\cdot10^{-9}$ &  hydrodynamic friction \cite{stokes1851effect}: $2\cdot6\pi\eta_c r_r$ - the receptor is assumed to be spherical
\vspace{0.1cm}\\
$L_1$ & m & $5\cdot10^{-6}$ & average of the observed jumps in the experimental EVs trajectories
\vspace{0.1cm}\\
$L_2$ & m & $6.9\cdot10^{-7}$ & distance of neuronal receptors estimated to be about the EV size to allow for rolling movement. \\
\bottomrule
    \end{tabular}
    \caption{List of model constants and  biophysical parameters values that have been estimated either by direct experimental measurements or by well-established physical formulas.}
    \label{table:physical_parameters}
\end{table}

\begin{table}
\centering 
    \begin{tabular}{l p{8cm} r r}
    \toprule
     \textbf{Parameter} &  \textbf{Description} & \textbf{Value} & \textbf{Unit }
\\
 \midrule
 
$k$  & spring stiffness &  $8\cdot10^{-9}$ & N/m
 \vspace{0.1cm} \\
$h_1$ & height of saw-tooth potential $V_1$ & $1.7128\cdot10^{-14}$ & J
\vspace{0.1cm} \\
$h_2$ & height of saw-tooth potential $V_2$ & $2.141\cdot10^{-14}$ & J
\vspace{0.1cm} \\
$\alpha$ & symmetry factor of the potential $V_1$ & $0.2$ & [-]
\vspace{0.1cm} \\
$\pi^\text{off}_1$ & passive unbinding rate  $(i.e. V^1_{S_1 = 1\rightarrow S_1 = 0})$  &  $500$ & Hz
\vspace{0.1cm} \\
$\pi^\text{on}_1$ & passive binding rate $(i.e. V^1_{S_1 = 0\rightarrow S_1 = 1})$ & $180$ & Hz
\vspace{0.1cm} \\
$\pi^\text{off}_2$ & active unbinding rate  $(i.e. V^2_{S_2 = 1\rightarrow S_2 = 0})$  &  $10$ & Hz
\vspace{0.1cm} \\
 $\pi^\text{on}_2$ & active binding rate  $(i.e. V^2_{S_2 = 0\rightarrow S_2 = 1})$ & $600$ & Hz
\vspace{0.1cm} \\
$\eta_\text{PA}$ & $\text{Passive}\rightarrow \text{Active}$ transition rate & $100$ & Hz
\vspace{0.1cm} \\
$\eta_\text{AP}$ & $\text{Active} \rightarrow \text{Passive}$ transition rate &  $200$ & Hz\\
\bottomrule
\end{tabular}
\caption{List of model parameters that have been calibrated to match available experimental data.}
    \label{table:fitted_parameters}
\end{table}

The system is simulated over $N_\text{runs}=200$ independent runs with a time-step $\Delta t = 10^{-4}$
for $N_T=12\cdot10^6$ steps, in order to reproduce the behavior of the EVs for 20 min, i.e. the average length of the experimental movies. The numerical displacements are then sampled with the same acquisition frequency of the experimental data (2 Hz).

\subsubsection*{Numerical vs \textit{in-vitro} results}

\begin{figure}
\centering
    \begin{tabular}{c c c}
        Ctrl & CytoD-EV & CytoD-HN \\
        \includegraphics[width=0.33\linewidth, trim=10 0 45 40,clip]{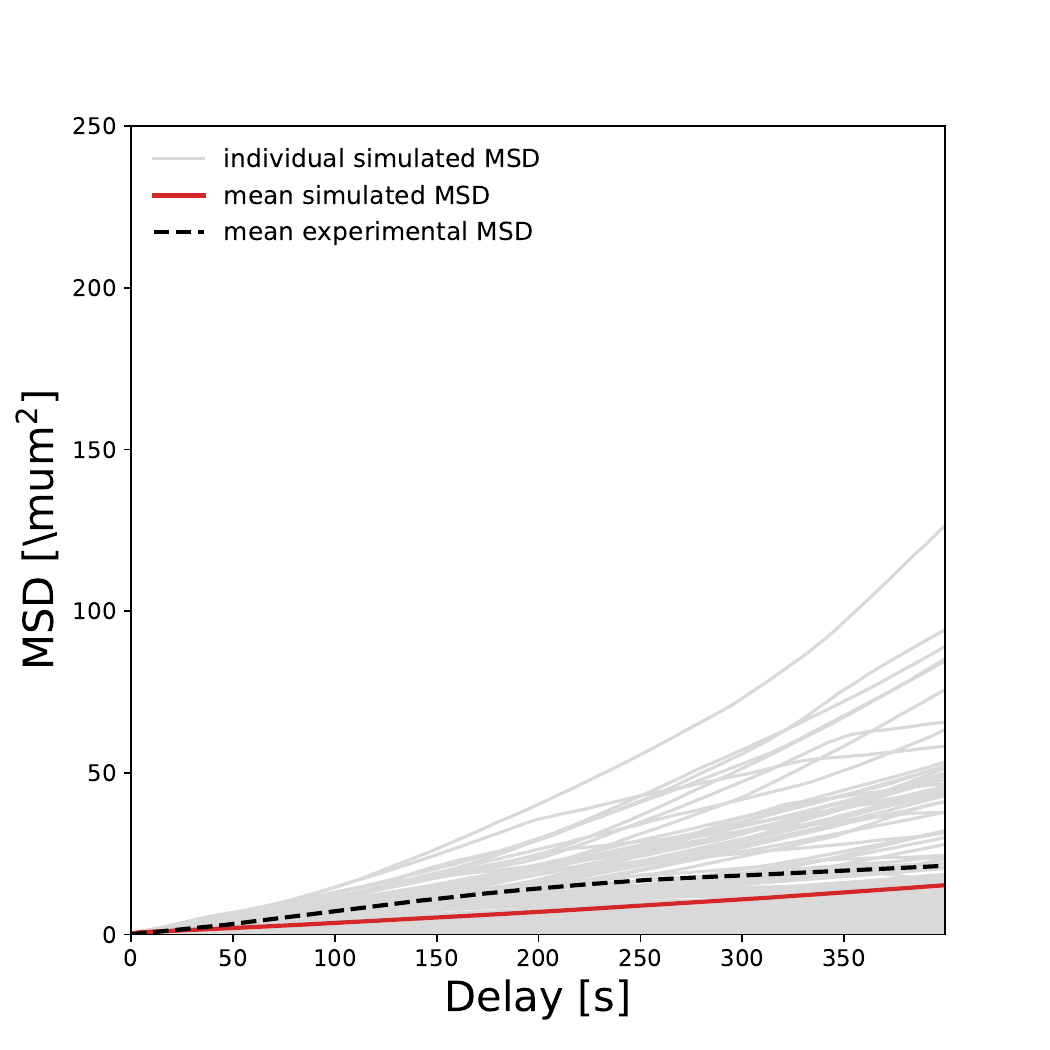}%
        &  \includegraphics[width=0.33\linewidth, trim=10 0 45 40, clip]{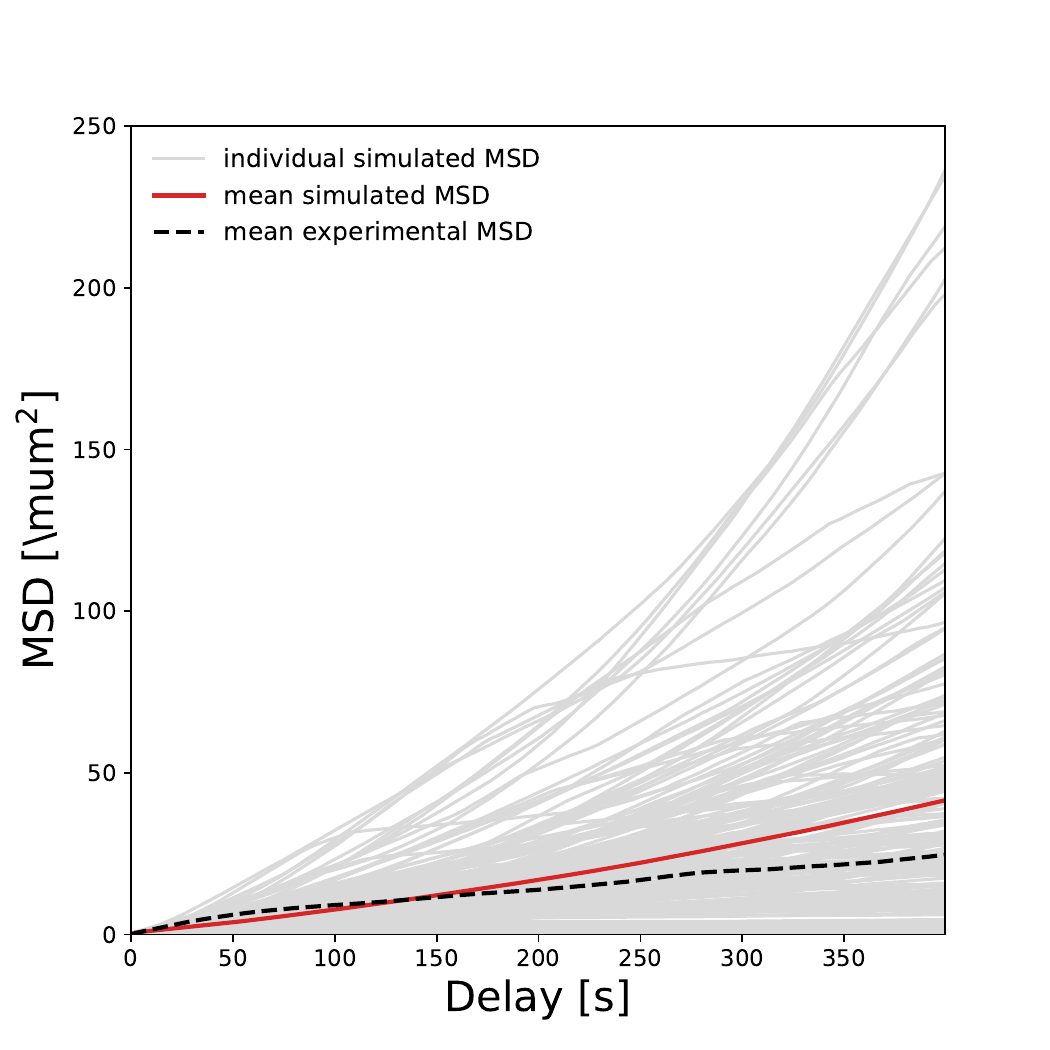} %
        & \includegraphics[width=0.33\linewidth, trim=10 0 45 40,clip]{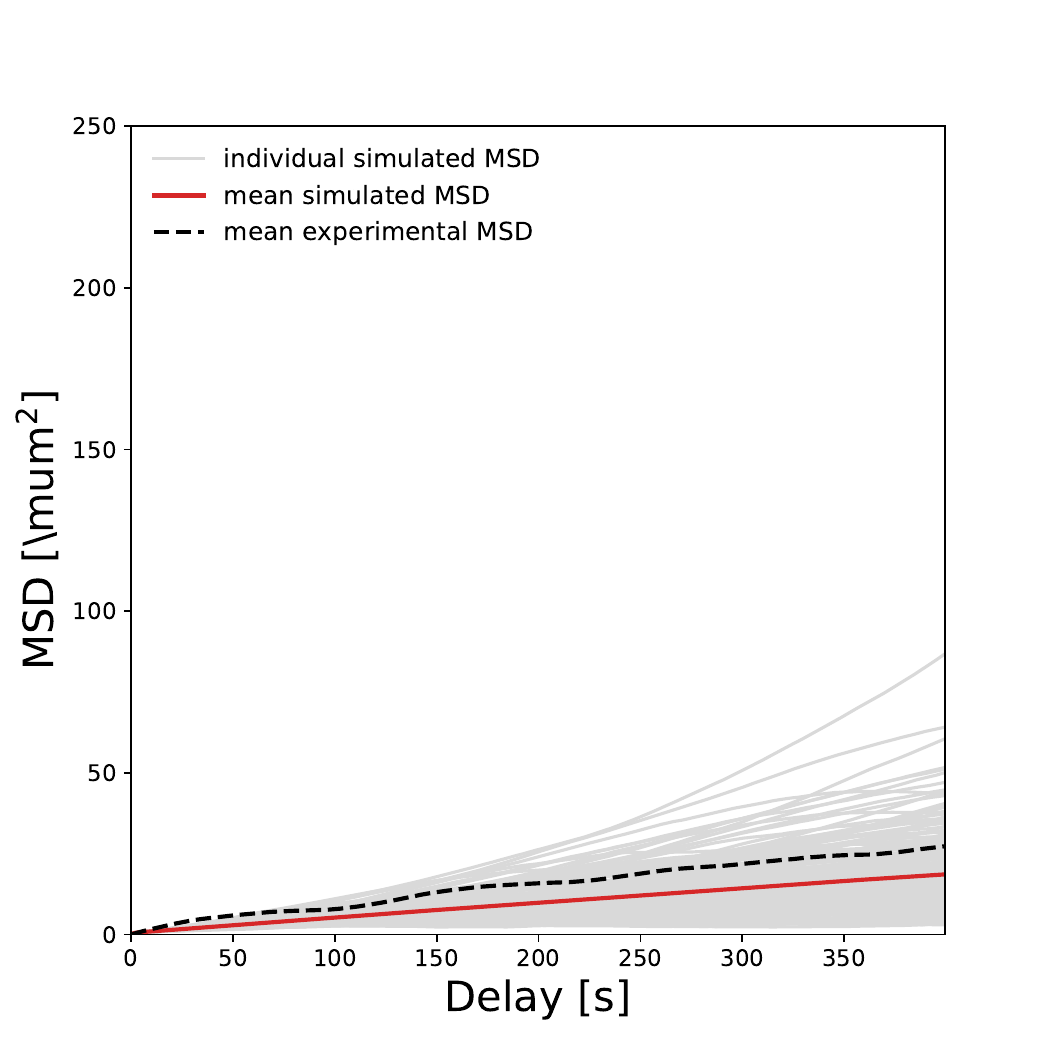} %
    \end{tabular}
        \caption{Comparison between the averaged Mean Square Displacement (MSD) obtained from the experimental data (black dashed line) and from the numerical simulations (red continuous line) for the \textit{Ctrl} set (left), the \textit{CytoD-EV} set (center) and the \textit{CytoD-HN} set (right). The gray lines represent the MSD relative to each individual numerical simulation.}
    \label{fig:msd_confronto}
\end{figure}

From a skewness test performed on simulated data, we found that, in agreement with the experimental data, the only set not showing a drifting component in the displacements is the CytoD-HN set (Ctrl \textit{p-value} = 0.0001, CytoD-EV \textit{p-value}  0.04, CytoD-HN \textit{p-value} = 0.57, with $N_{runs}$ = 200 for each set).
This feature also weakly emerges from the numerical Mean Squared Displacement (MSD) curves reported in Fig. \ref{fig:msd_confronto}. Indeed, we observe an MSD close to super-diffusive (indicating the presence of a drift) for the CytoD-EV set, close to linear for the Ctrl set (in accordance with the higher p-value of the test), whereas the CytoD-HN set MSD seems to indicate a sub-diffusive dynamics.
Although the limited number of experimental samples (Ctrl: $ n = 15$, CytoD-EV: $n = 13$, CytoD-HN: $n = 14$) makes difficult to identify clear trends of the experimental data, the numerical and experimental curves show good overall qualitative agreement. However, there exist some discrepancies in magnitude, suggesting that additional factors may influence the transport dynamics.

\begin{figure}
\centering
    \begin{tabular}{c c c}
        Ctrl & CytoD-EV & CytoD-HN \\
        \includegraphics[width=0.33\linewidth, trim=5 0 45 40,clip]{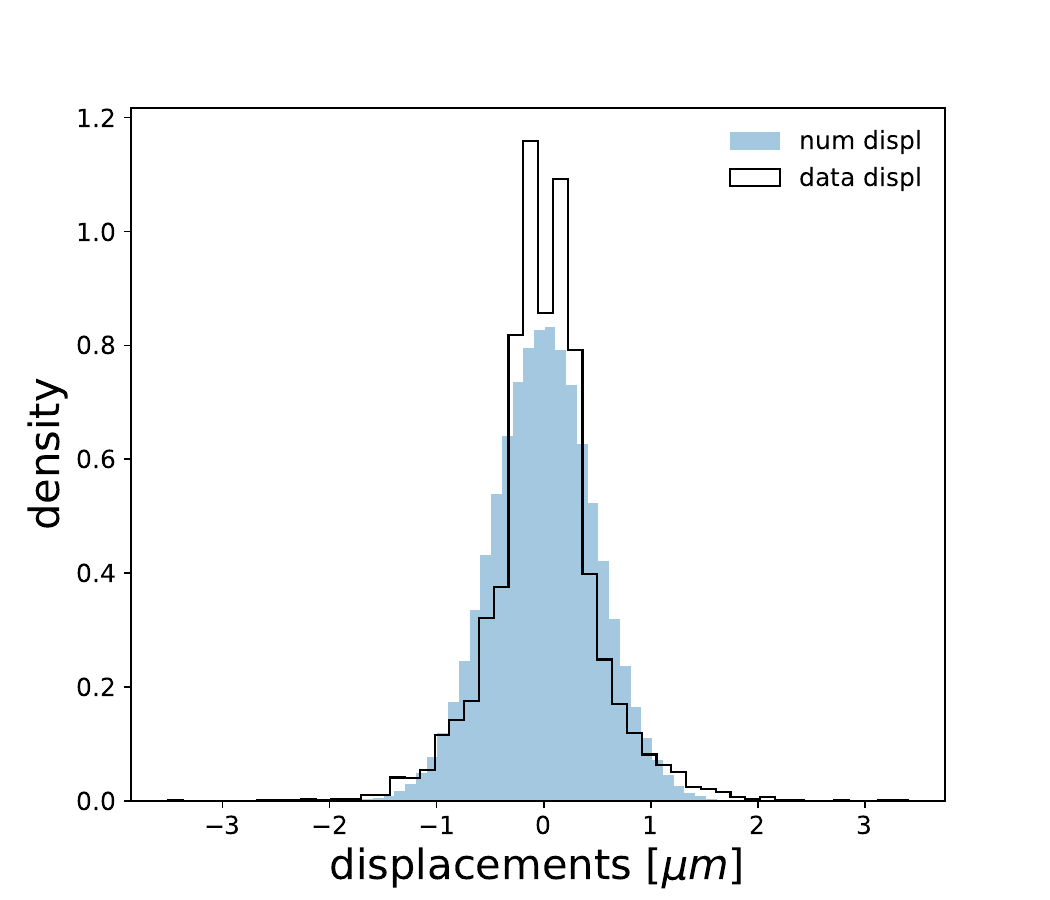}%
        & \includegraphics[width=0.33\linewidth, trim=5 0 45 40, clip]{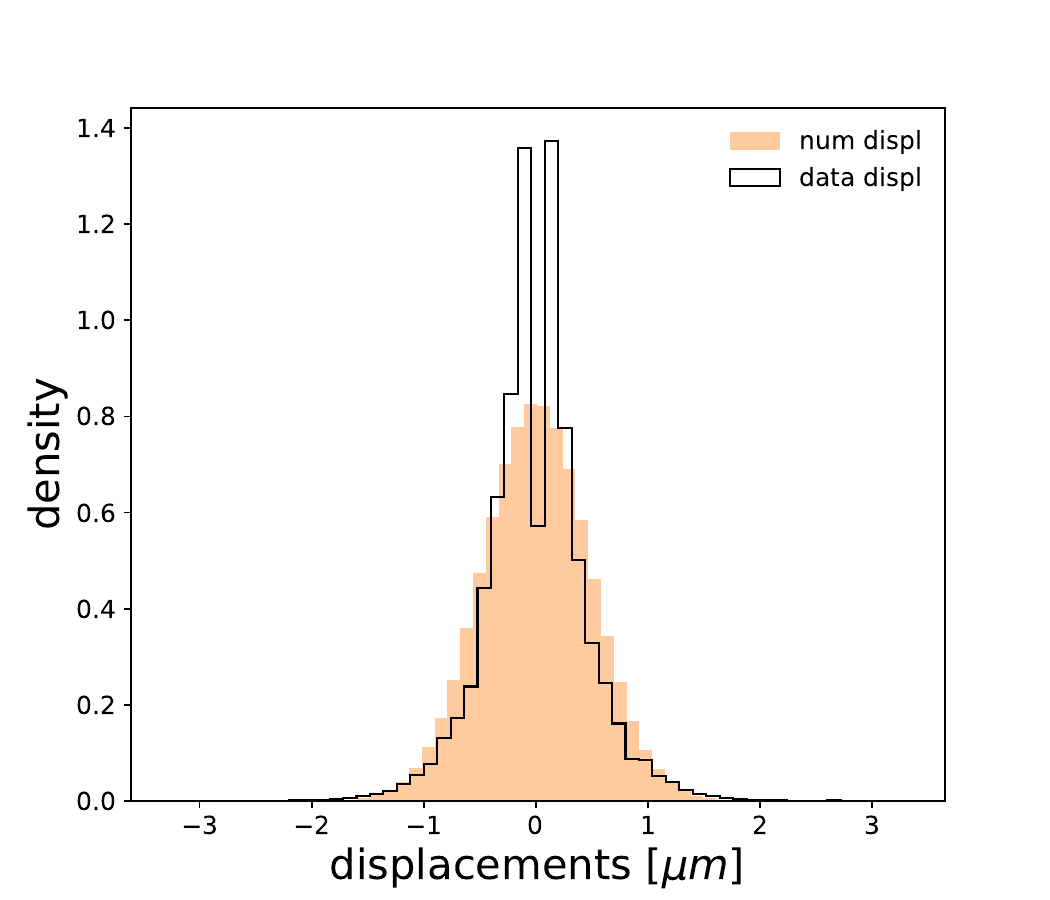} %
        & \includegraphics[width=0.33\linewidth, trim=10 0 45 40,clip]{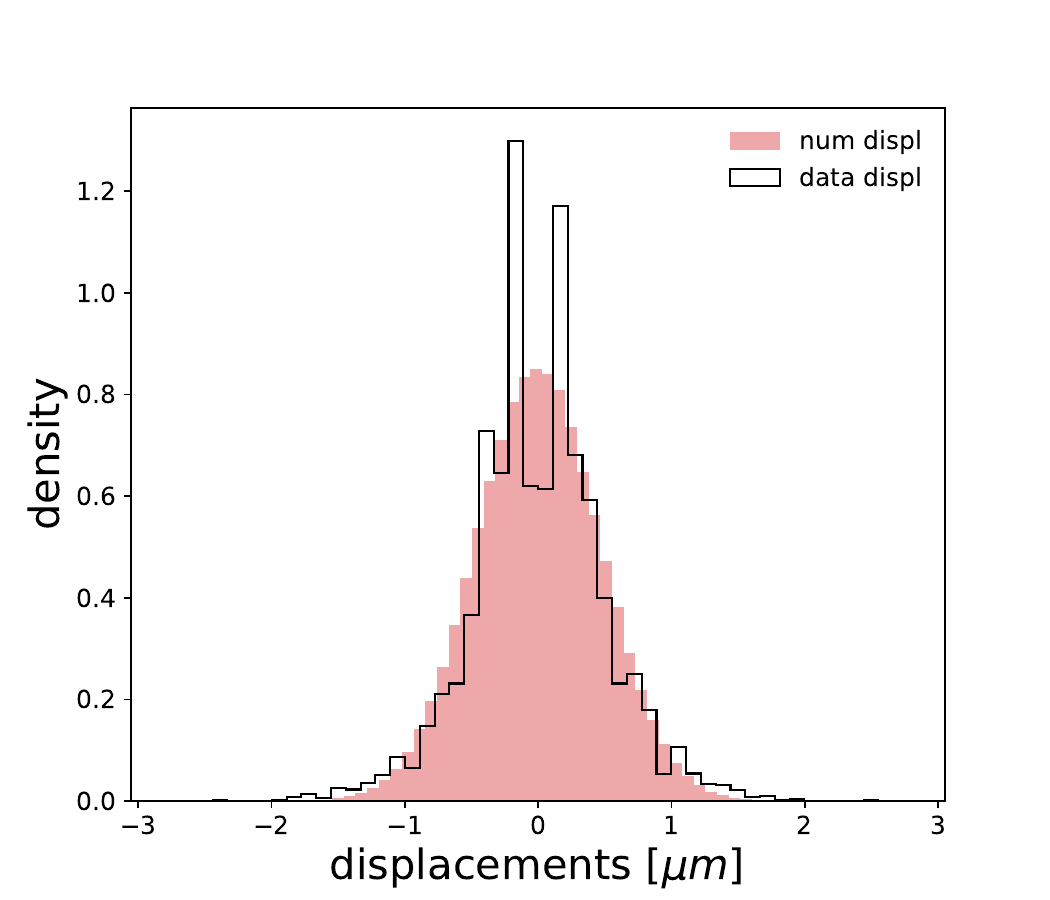} %
    \end{tabular}
        \caption{Comparison between the histogram of the non-zero displacement (black continuous profile) and the one of the numerical displacements (shaded areas) for the \textit{Ctrl} set (left), the \textit{CytoD-EV} set (center) and the \textit{CytoD-HN} set (right).}
    \label{fig:hist_confronto}
\end{figure}

The histograms reported in Fig. \ref{fig:hist_confronto} show a good agreement of the numerical displacement distribution with that of the experimental data, both in terms of the mean and variance of the distributions.
Instead, the discrepancy between the unimodal shape of the numerical displacement distribution and the bimodal shape of the experimental data can be attributed to the combined effect of experimental error in data acquisition and the filtering of null experimental displacements.

\begin{figure}
    \centering
    \includegraphics[width=0.325\linewidth, trim=5 0 30 0,clip]{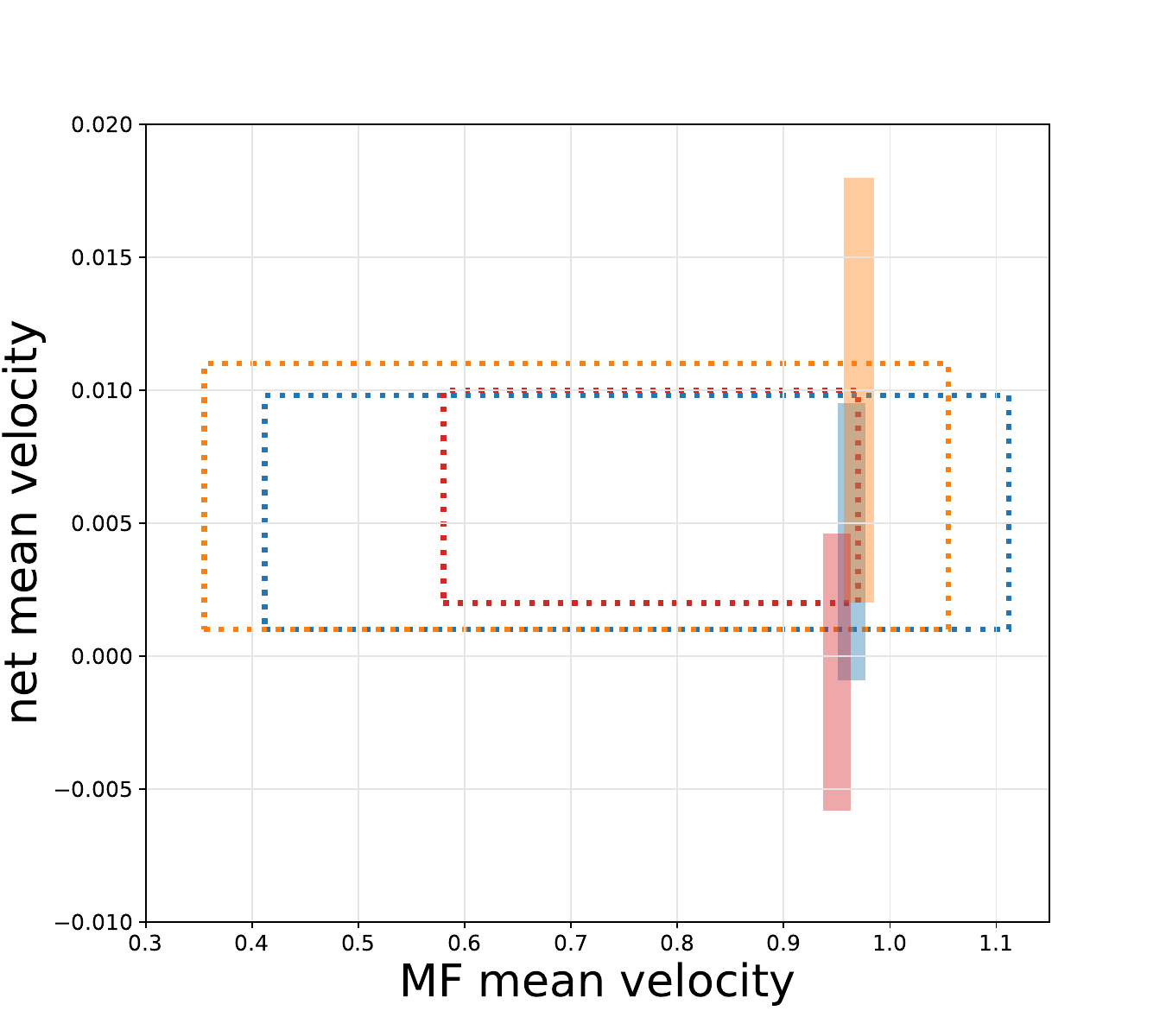} %
    \includegraphics[width=0.325\linewidth, trim=5 0 30 0,clip]{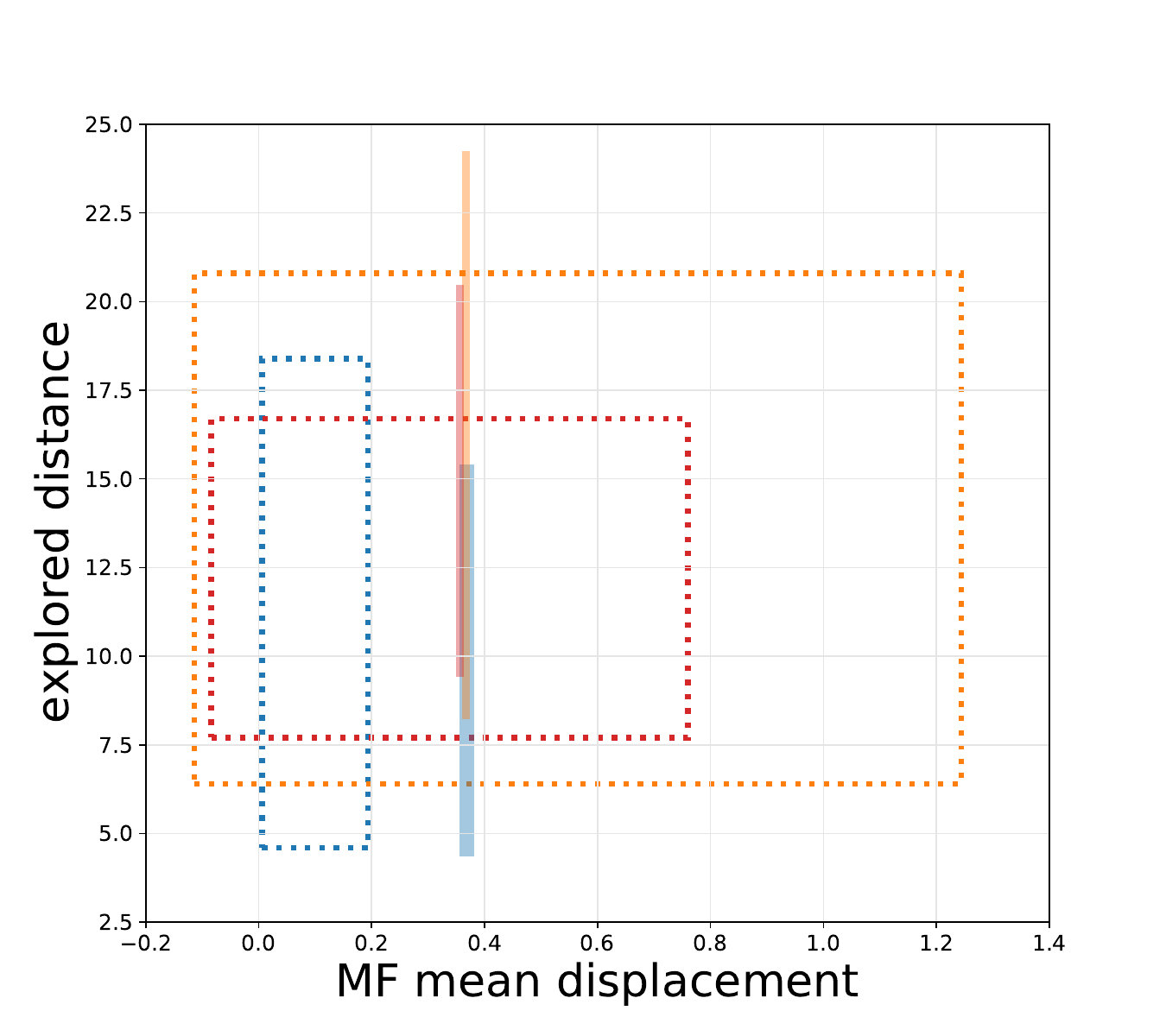} %
    \includegraphics[width=0.325\linewidth, trim=5 0 30 0,clip]{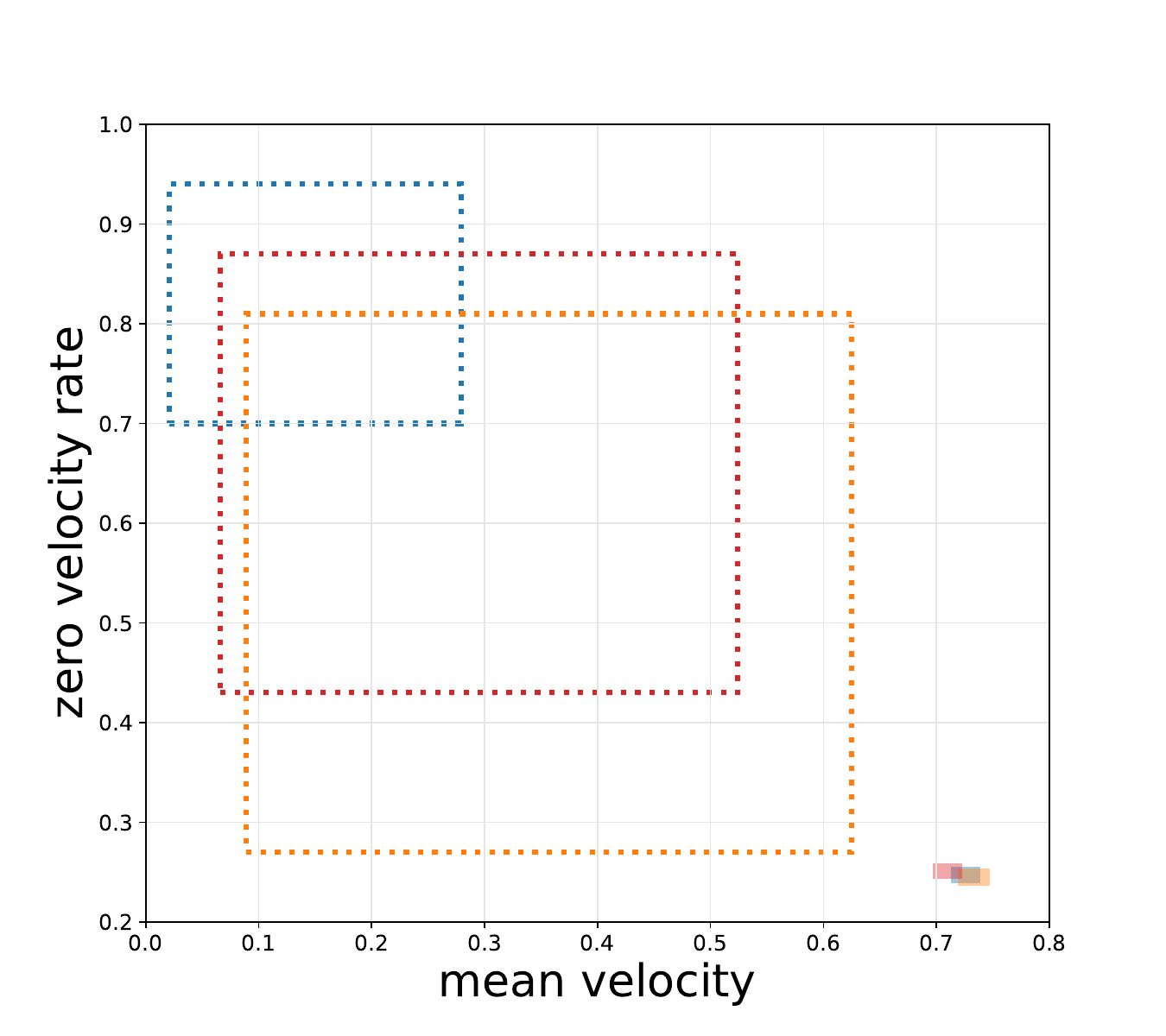} \\
    \includegraphics[width=0.34\linewidth]{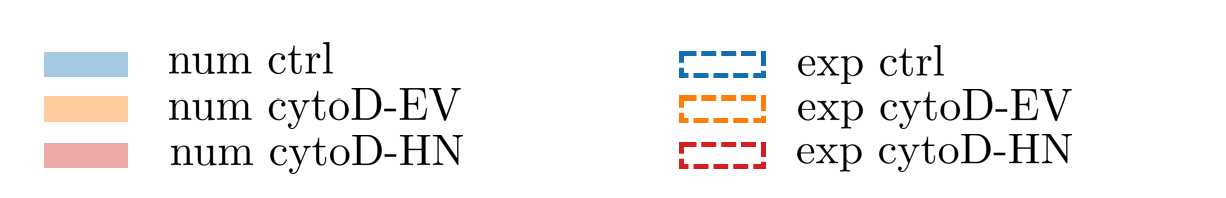}
    \caption{Comparison between the experimental (dashed) and numerical (shaded) ranges for the main kinematic indicators. Results for the \textit{Ctrl} set (blue), for the \textit{CytoD-EV} set (orange) and for the \textit{CytoD-HN} (red) are reported.
    (Left) \textit{MF mean velocity - net mean velocity} plane. (Center) \textit{MF mean displacement - explored distance} plane. (Right) \textit{mean velocity - zero velocity rate} plane. In all the cases, the rectangles are centered on the mean values and their width and height correspond to twice the standard deviation of the considered indicators.}
    \label{fig:intervalli_confronto}
\end{figure}

Finally, in Fig. \ref{fig:intervalli_confronto} we show the comparison between a subset of the kinematic indicators obtained from experimental data and numerical simulations. With the exception of \textit{mean velocity} and \textit{zero velocity rate}, there is a discrete level of overlap in the confidence intervals (i.e. each rectangle edge, obtained as mean $\pm$ variance) for the indicators in all the three sets.
In particular, while \textit{net mean velocity} and \textit{explored distance} exhibit a near complete overlap of the confidence intervals, the numerical ranges of \textit{MF mean velocity} and \textit{MF mean displacement} are noticeably narrower than the corresponding experimental ranges suggesting that for these indicators the data variability is not fully captured by the model.

\subsubsection*{Sensitivity analysis to parameter variations}

\begin{figure}[H]
    \centering
    {
    \includegraphics[width=0.45\textwidth]{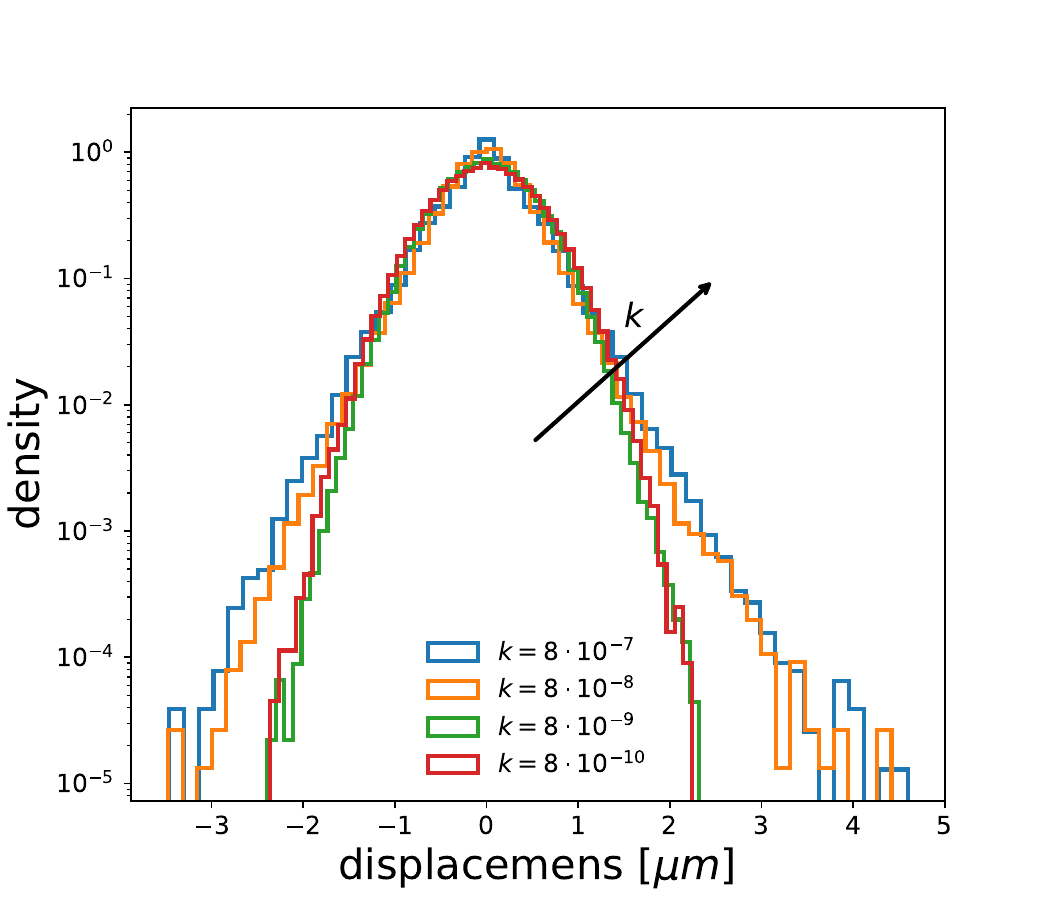}%
    \includegraphics[width=0.45\textwidth]{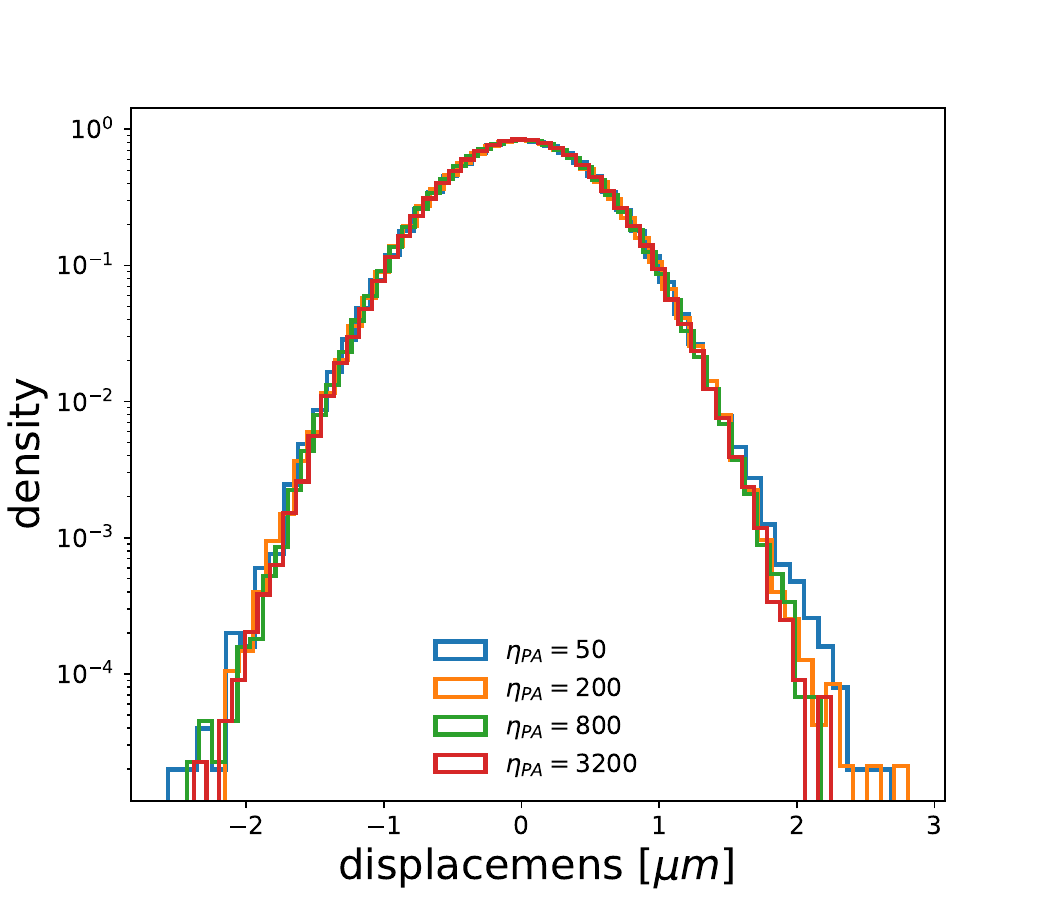}
    \includegraphics[width=0.45\textwidth]{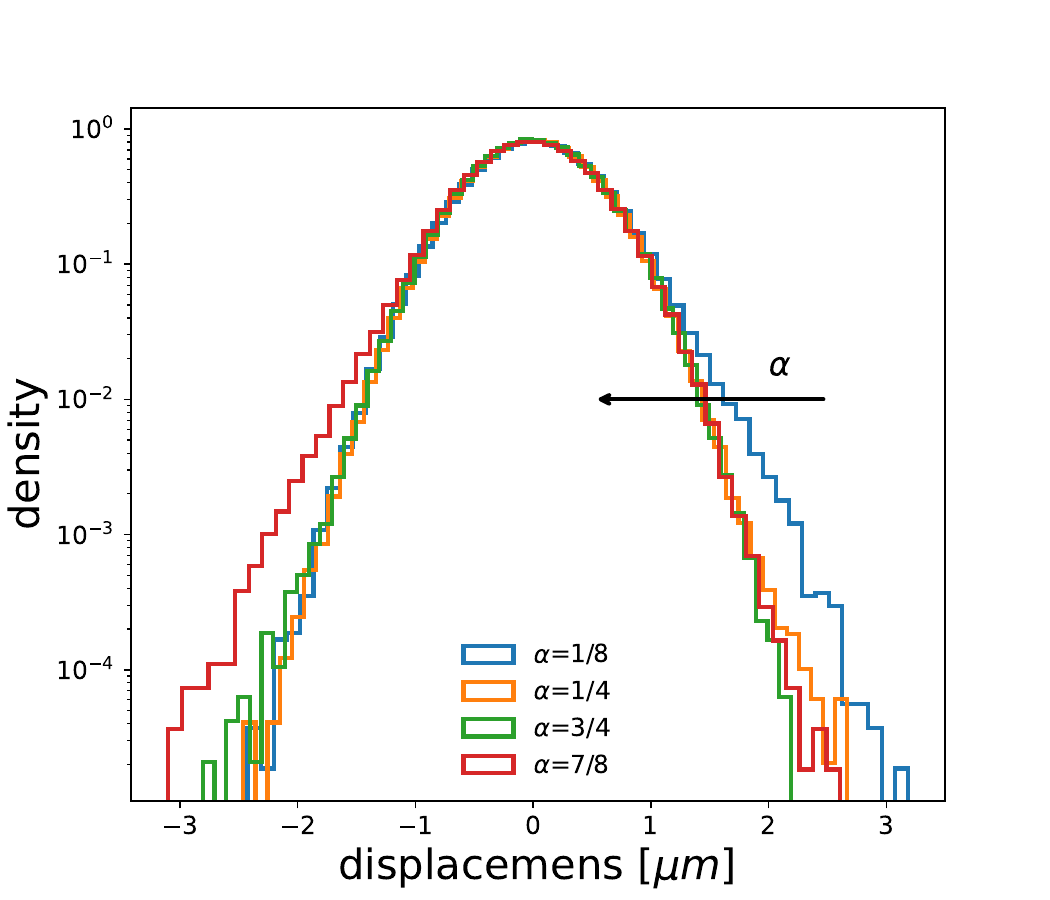}%
    \includegraphics[width=0.45\textwidth]{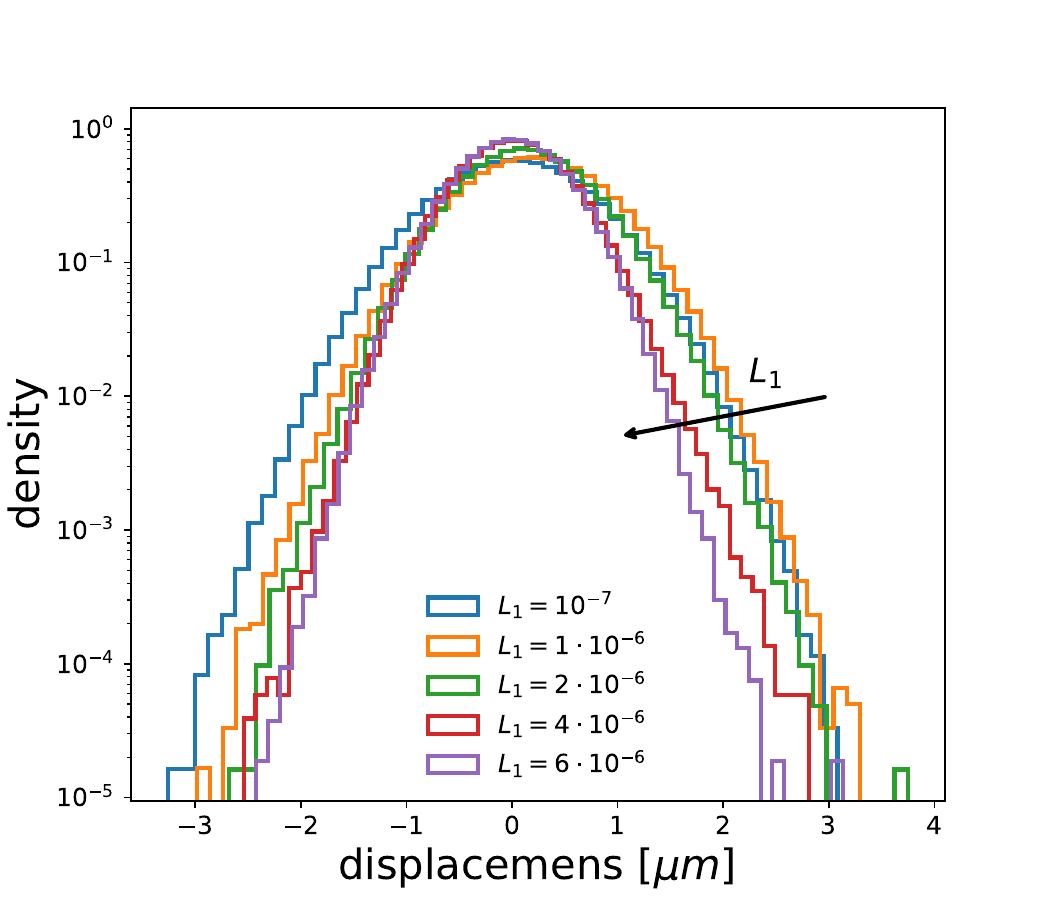}
    \includegraphics[width=0.45\textwidth]{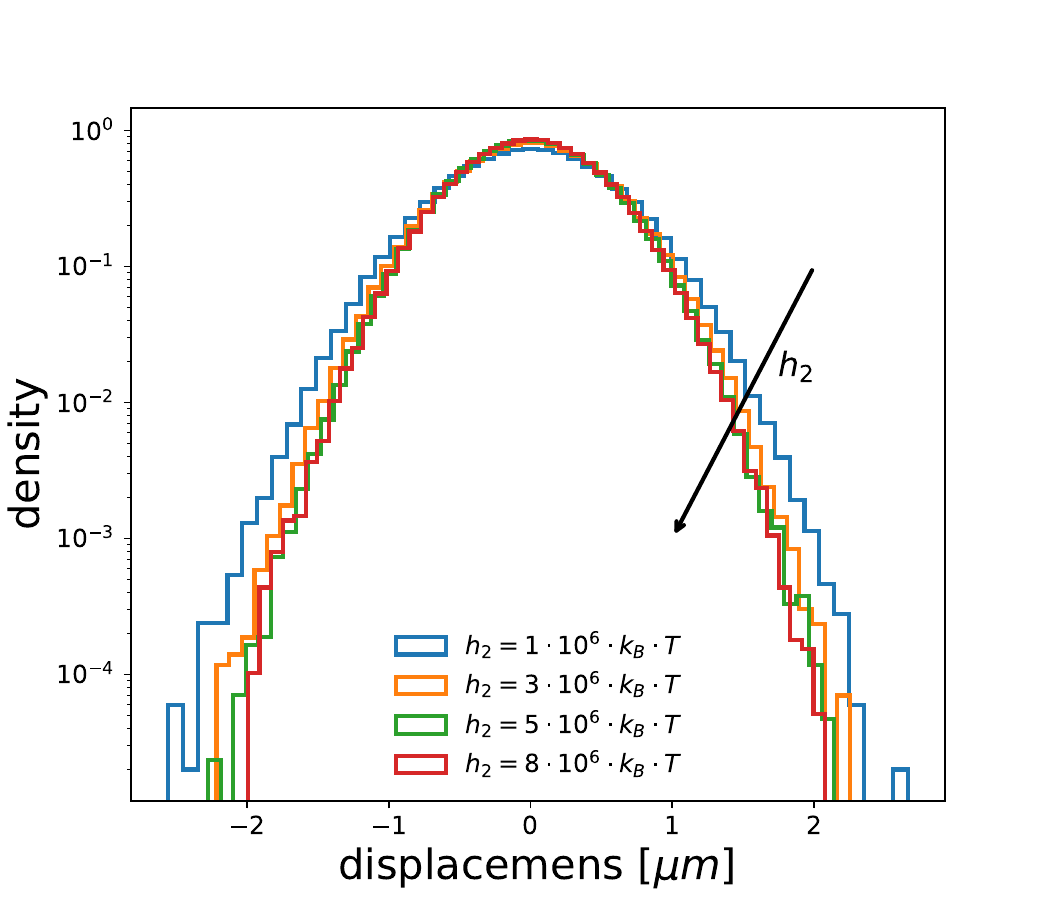}%
    \includegraphics[width=0.45\textwidth]{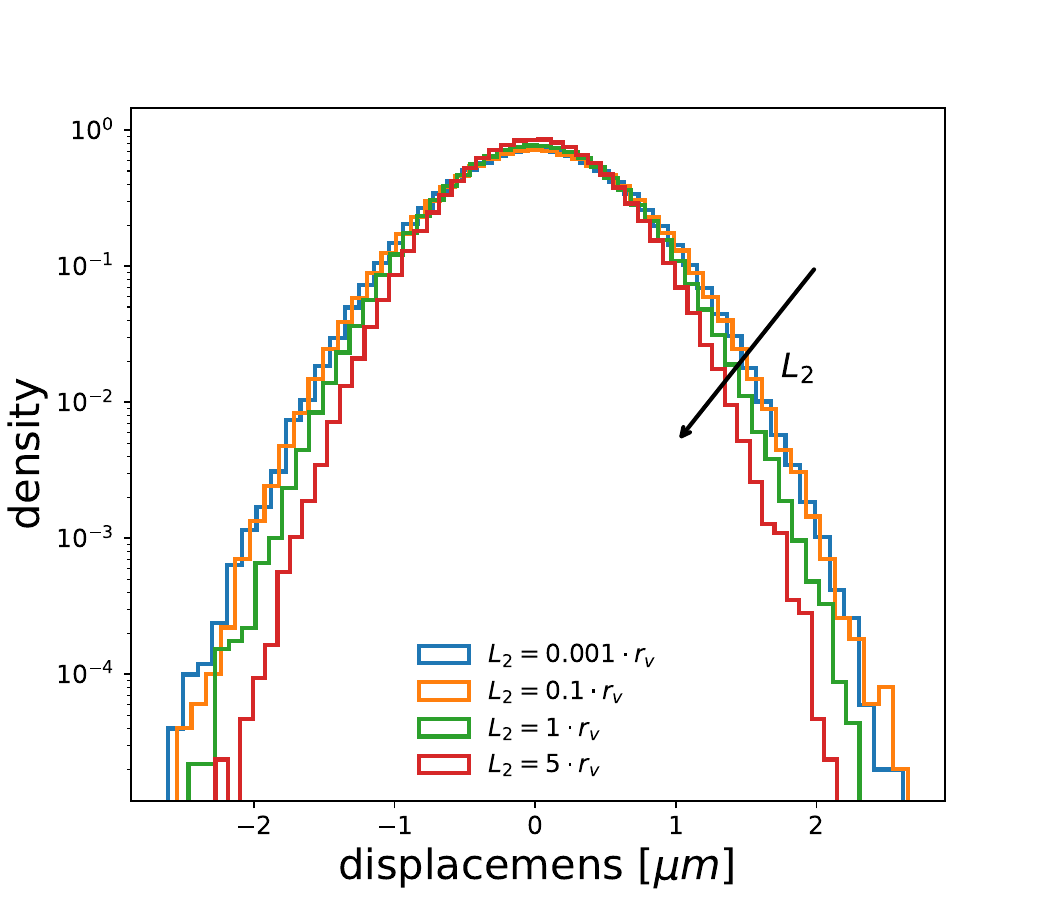}
    }
    \caption{Sensitivity analysis performed varying a subset of the model parameters: the spring stifness $k$ (top left), the passive to active transition rate $\eta_{PA}$ (top right) in the \textit{Ctrl} scenario; the symmetry parameter $\alpha$ (center left) and the period $L_1$ of the asymmetric saw-tooth potential $V_1$ (center right) in the \textit{CytoD-EV} scenario; the height $h_2$ (bottom left) and the period $L_2$ (bottom right) of the symmetric saw-tooth potential $V_2$ in the \textit{CytoD-HN} scenario. Where a clear trend is present, black arrows indicate the increase of the parameter.}
    \label{fig:sensitivity_ctrl}
\end{figure}

In this Section we perform a sensitivity analysis of the proposed mathematical model, wherein the parameter values are varied one at a time while the remaining parameters are held constant. The outcomes of this analysis are reported in Fig. \ref{fig:sensitivity_ctrl}.

We first tested the effect of varying the spring stiffness $k$ and the transition rate from the passive to the active state $\eta_\text{PA}$ in the Ctrl scenario. We notice that the stiffening of the bond between the Ev and the axon produce larger displacements and this could be due to the high ability of the receptor to recall the vesicle against thermal fluctuations, see Fig. \ref{fig:sensitivity_ctrl} (\textit{top left}).
Unbalancing the system dynamics in favor of the active transport state over the passive transport (i.e. increasing $\eta_\text{PA}$), makes the slight drift generated by $V_1$ gradually disappear, see Fig. \ref{fig:sensitivity_ctrl} (\textit{top right}).

We then tuned the symmetry parameter $\alpha$ and the period $L_1$ associated to the flashing ratchet potential $V_1$ in the CytoD-EV scenario.
We observe that for $\alpha<1/2$, the histograms are shifted to the right, meaning that the EV transport is biased towards positive values. On the contrary, for $\alpha>1/2$ the drift shifts towards negative values. 
Overall, as expected, the transport is  more efficient for extreme values of $\alpha$, see Fig. \ref{fig:sensitivity_ctrl} (\textit{center left}).
We then notice that the largest drift is attained for the lowest value of $L_1$, see Fig. \ref{fig:sensitivity_ctrl} (\textit{center right}).
We believe that it becomes easier with a shorter period for the particle to diffuse and jump to the next potential minimum subject to the flashing ratchet, while with a larger period the particle would have to diffuse over a much longer distance in order to attain the next minimum, significantly decreasing the probability of this event.

Finally, we explored the effect of tuning the height $h_2$ and the period $L_2$ of the symmetric saw-tooth potential $V_2$ in the CytoD-HN scenario.
As expected, the larger $h_2$ the lower the displacements, see Fig. \ref{fig:sensitivity_ctrl} (\textit{bottom left}). Indeed, this parameter represents the energy barrier that the EV has to cross in order to actively move to the next receptor. 
We observe the same behavior rising the period $L_2$, i.e. lowering the density of receptors on the neuron surface, for which the same considerations made for $L_1$, see Fig. \ref{fig:sensitivity_ctrl} (\textit{bottom right}).

\section{Discussion and conclusions}
\label{sec::discussion}

This study investigated the movement of EVs on neuronal surfaces, analyzing three experimental datasets.

The experimental procedure involved isolating EVs from astrocytes and placing them to neuron surfaces, with or without Cytochalasin D treatment. Live imaging of EVs allowed for the tracking of their trajectories, which were then analyzed using custom \texttt{MATLAB} scripts. Statistical analysis using Python libraries provided insights into the kinematic indicators of EV movement, such as mean velocity, zero-velocity rate, and path length. These indicators highlighted significant differences between the Ctrl and CytoD-treated datasets, with the Ctrl group exhibiting lower mobility and higher zero-velocity rates. The datasets were categorized as follows: Ctrl dataset (control, no treatment), CytoD-HN dataset (neurons treated with Cytochalasin D), and CytoD-EV dataset (EVs treated with Cytochalasin D). This categorization allowed for a comparative analysis of how actin polymerization inhibition affects EV transport.

The mathematical model we proposed aims at explaining
the experimental data by incorporating two distinct transport mechanisms: passive transport and active transport. In the Ctrl dataset, EVs exhibited more constrained diffusion, suggesting a predominance of passive transport mediated by cytoskeletal activity, similarly to the CytoD-EV case where this aspect is intensified. This passive transport was modeled using a flashing Brownian ratchet with an asymmetric sawtooth potential, which allowed us to replicate the directed movement observed experimentally. On the other hand, the active transport of EVs, the dominant mechanism in the CytoD-HN set, was indeed modeled through a flashing Brownian ratchet model with a symmetric periodic potential. 
The model parameters were calibrated based on experimental measurements and well-established physical relations. Parameters such as the spring stiffness, the heights of the sawtooth potentials, and the transition rates between active and passive states were crucial in accurately capturing the dynamics of EV movement.

Overall, the inhibition of actin polymerization by Cytochalasin D significantly impacted the EVs' ability to move along the neuron. 
In the CytoD-EV dataset, the EVs still retained a directed transport component, indicating that even when EVs were treated with CytoD, they could still exhibit passive movement movement due to cytoskeletal actin filament rearrangements. 
Conversely, the CytoD-HN dataset showed no drifting component, consistent with a purely active transport mechanism where cytoskeletal dynamics were disrupted and EVs can move by simply rolling on the neuron surface leveraging nearby receptors.

Our primary findings indicate that EVs exhibit distinct mobility patterns depending on the treatment applied. Specifically, EVs in the Ctrl group demonstrated reduced mobility compared to those in the CytoD-treated groups, as evidenced by a higher percentage of zero-velocity occurrences and lower values of path length and mean velocity.

The model's plausibility was validated by comparing numerical simulations with experimental data. A skewness test on simulated data showed agreement with experimental observations, where only the CytoD-HN set lacked a drifting component in the displacements. The mean squared displacement (MSD) curves further supported these findings, showing a  super-diffusive trend for the CytoD-EV set, possibly indicative of a drift, while the CytoD-HN set exhibited a sub-linear MSD, reflecting purely diffusive behavior. Additionally, histograms of displacement distributions from numerical simulations closely matched those from experimental data, 
enforcing the model's capability of capturing the underlying transport dynamics. Future improvements  in data analysis will concern clustering EV trajectories based on kinematic indicators to identify subpopulations with distinct transport dynamics. Applying unsupervised learning techniques could help to better distinguish between passive and active transport regimes.

Our sensitivity analysis revealed that varying model parameters, such as spring stiffness and the transition rates between active and passive states, significantly affected EV transport characteristics. Increasing the spring stiffness resulted in larger displacements, suggesting a stronger ability of the receptor to pull the EV against thermal fluctuations. Adjusting the transition rates indicated that an imbalance favoring active transport diminished the slight drift generated by the asymmetric sawtooth potential.
Furthermore, a decrease in the density of receptors on the neuron surface, as well as an increase in the intensity of the bond, impairs Ev motility.

However, the model has certain limitations. Firstly, it assumes that the EVs and receptors behave as spherical particles, which may oversimplify the actual biological structures.
Additionally, the model does neither account for the EVs shape deformation, due to the presence of actin filaments in their lumen, nor for potential interactions with other cellular components or environmental factors that could influence EVs movement.
The Markov processes governing state transitions are also simplified and may not fully capture the complexity of the biological mechanisms involved. Furthermore, the calibration of model parameters is based on a low number of available experimental samples, which may introduce biases or inaccuracies due to measurement limitations.
Future research directions include further refinement of the mathematical model to gain a more detailed understanding of the molecular mechanisms underlying EV transport. For example, incorporating the actin-based mechanism regulating EV shape could enhance the model’s predictions, since it affects both the rotational dynamics of EV movement and the elasticity of the binding.

Overall, while our model effectively captures key features of EV transport, including distinct motility patterns across experimental conditions, we acknowledge that the limited sample size and inherent variability in single-particle trajectories introduce statistical uncertainty. Although trends in MSD and displacement distributions are qualitatively reproduced, some discrepancies in magnitude remain, suggesting that additional factors, such as heterogeneity in EV properties or local receptor density, may influence transport dynamics. Further experimental validation with larger datasets would be necessary to fully confirm the proposed mechanistic interpretations and to reduce data variability. Nevertheless, our findings provide a plausible framework for understanding the interplay between passive and active transport mechanisms in EV motion.
However, although CytoD treatment disrupts the actin cytoskeleton, it does not directly address whether EV transport is an ATP-dependent, actively driven process. Further investigation is needed to determine whether the observed movement is truly super-diffusive or influenced by energy-driven molecular motors. Future experiments using inhibitors such as blebbistatin, which specifically targets myosin-II ATPase activity, could help clarify whether different ATP-dependent mechanisms contribute to EV transport on the neuronal surface. Moreover, we aim at  measuring the protein abundance of PrP with a single particle analysis to estimate more precisely the protein abundance on their surface. 
Our long term goal is to refine this modelling framework to specific pathological conditions, such as neurodegenerative diseases, possibly providing insights into how variations in EV transport mechanisms contribute to the spread of misfolded proteins, potentially identifying and thus modeling new therapeutic targets.

In summary, our work presents a new modelling and computational framework for understanding EV behavior and transport mechanisms, with significant implications for research into neurodegenerative diseases and other pathological conditions. The combination of experimental data and mathematical modeling has provided valuable insights into the biophysical processes governing EV movement on neuronal surfaces, paving the way for future advancements in understanding the biophyisical laws underlying the prion protein-mediated transport of EVs on the neuron surface, and its key role for activation \cite{lombardi2021role} and evolution of brain diseases \cite{ruan2020p2rx7}.

\section*{Acknowledgement}
GP, PC and ST have been partially supported by MUR, PRIN Research Projects 2020F3NCPX and grant Dipartimento di Eccellenza 2023-2027.
GP has been partially supported by GNFM – INdAM through the
program Progetto Giovani 2023. GP, PC and ST gratefully acknowledges the support from the Istituto Nazionale di Alta Matematica (INdAM) and the Gruppo Nazionale per la Fisica Matematica (GNFM).
We are thankful to Micol Boriotti for her earlier contribution to the model development.

%%%%%%%%%%%%%%%%%%%%%%
\newpage

%\nocite{*} 
%\phantomsection 
%\addcontentsline{toc}{chapter}{\numberline{}\bibname}\printbibliography

\bibliographystyle{plain} % Scegli lo stile della bibliografia
\bibliography{evs} % Nome del file .bib (senza estensione)
\end{document}

\title{{\bf Supplementary Material}: 
\newline Modeling the prion protein-mediated transport of extracellular vesicles on the neuron surface}
\author{G.~Pozzi$^{1}$,  G.~Mazzilli$^{2}$, G.~D'Arrigo$^{3}$, C.~Verderio$^{3}$, G.~Legname$^{4}$, \\S.~Turzi$^{5}$, P.~Ciarletta$^{2}$\footnote{Corresponding author. E-mail: pasquale.ciarletta@polimi.it} \\
	$^{1}$ DISMA, Politecnico di Torino, corso Duca degli Abruzzi 24, 10129 Torino, Italy\\
    $^{2}$ MOX Laboratory, Dipartimento di Matematica, Politecnico di Milano,\\ piazza Leonardo da Vinci 32, 20133 Milano, Italy. \\
    $^{3}$ Institute of Neuroscience of Milan, CNR National Research Council of Italy, Vedano al Lambro, 20854, Italy.\\
    $^{4}$ Department of Neuroscience, Scuola Internazionale Superiore di Studi Avanzati (SISSA), Trieste, Italy.\\
    $^{5}$ Dipartimento di Matematica, Politecnico di Milano, piazza Leonardo da Vinci 32, 20133 Milano, Italy.\\
    }

\maketitle
\date{}

\section{Density estimate of proteins and neurite receptors}
\label{SI::density_estimate}
In this Section we provide a temptative extimation of the density of proteins proteins per EV and neurite receptors to support the feasibility of the proposed transport mechanism involving PrP interactions with neuronal receptors.

To this purpose, we observe that EVs, particularly exosomes, exhibit a log-normal diameter distribution with a mode around 60–80 nm, sometimes extending up to 150 nm. Given their lipid bilayer structure (~4 nm thick), the available surface area for protein binding can be estimated as A=4$\pi$R$^2$, which for an 80 nm vesicle (R = 40 nm) results in approximately 2x10$^4$ nm$^2$. Typical membrane proteins occupy ~10 nm$^2$ per molecule, leading to a theoretical maximum packing density of ~2000 proteins per vesicle; however, steric hindrance and diffusion constraints reduce this to 400–1000 proteins per EV in physiological conditions \cite{thery2018minimal}. Neuronal surface receptor densities typically range from 100 to 1000 molecules/µm$^2$, with an average estimate of 500 receptors/µm$^2$, suggesting that for a 100-nm vesicle (~0.01 µm$^2$ contact area), about 5 receptors could be available for simultaneous binding \cite{sudhof2008neuroligins}. Considering that an EV can present up to 1000 surface proteins, including PrP, and that ligand-receptor interactions typically exert forces in the range of 1-100 pN \cite{chothia1975structural}, it is feasible that a single EV could engage multiple PrP-receptor interactions simultaneously. Assuming at least 5–10 receptor-binding events per vesicle, the cumulative force (5 × 10 pN = 50 pN) is sufficient to drive a rolling or hopping motion, particularly when supported by fluctuations in the cytoskeleton and local membrane dynamics \cite{brown2005stochastic}. Our estimates thus demonstrate that the proposed PrP-mediated transport mechanism is compatible with the physical constraints of EVs and neuronal membranes, even for the smallest EVs (~60 nm), supporting the plausibility of our model.

\section{Size distribution of astrocytes-derived EVs}
\label{SI::size_distribution}
Cell supernatant of 7 × 106 astrocytes was partitioned in two aliquots, after clearing from cell debris (300 × g × 10 min twice). One aliquot was centrifuged at 10,000 × g for 30 min, the pellet resuspended with Izon EV reagent kit and analysed using an NP400 pore. The other aliquot was further centrifuged at 2,000 × g for 20 min, to eliminate bigger EVs, before being pelleted at 10,000 × g, resuspended and analysed using an NP200. Removal of bigger EVs limited NP200 pore obstruction. Analysis by NP400 pore was used to determine the concentration of EVs $>$ 200 nm, while by NP200 pore we evaluated $<$ 200 nm EV concentration Note that medium/large EVs ($>$ 200 nm) are $\sim$58$\%$ of EV population (n = 2507).

\begin{figure}
    \centering
    \includegraphics[width=0.5\linewidth]{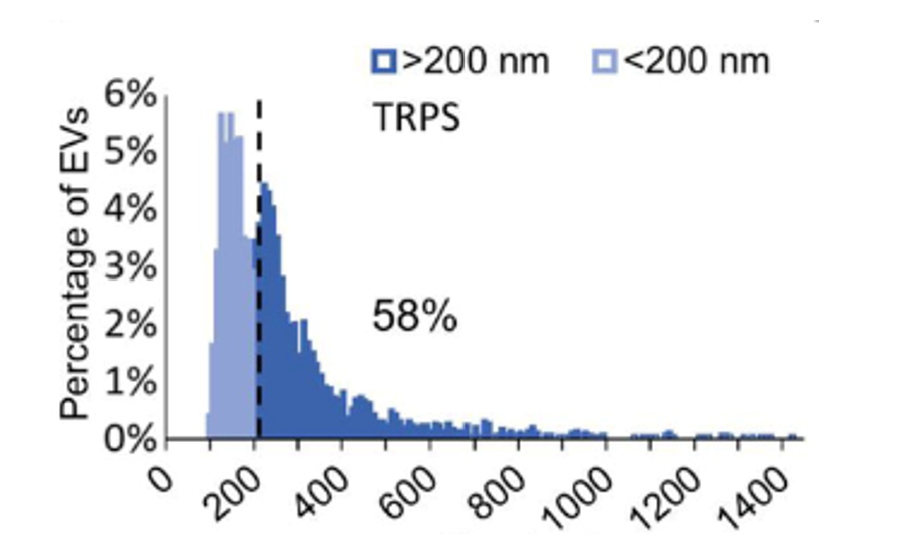}
    \caption{Size distribution of EVs in the 10,000 × g pellet according to TRPS. Image from ref. \cite{d2021astrocytes}.}
    \label{fig:size-distribution}
\end{figure}

\bibliographystyle{plain} % Scegli lo stile della bibliografia
\bibliography{evs} % Nome del file .bib (senza estensione)